\documentclass[aps,prb,floatfix,twocolumn,superscriptaddress,nopacs,longbibliography,epsf]{revtex4-2}

\usepackage[USenglish]{babel}
\usepackage{amsmath,amssymb,amsfonts}
\usepackage{epstopdf}
\usepackage{epsfig} 
\usepackage{graphicx}
\usepackage{subfigure}
\usepackage{import}
\usepackage{xcolor}
\usepackage{url}

\usepackage{siunitx}

\usepackage{tikz}
\usetikzlibrary{tikzmark}

\newlength{\mytextsize}

\usepackage{mathtools}

\allowdisplaybreaks

\usepackage[colorlinks=true,linkcolor=blue,citecolor=red]{hyperref}
\usepackage[final]{changes}

\makeatletter
\renewcommand\tableofcontents{\@starttoc{toc}}
\makeatother
\def\bcen{\begin{center}}
\def\ecen{\end{center}}

\def\a{\alpha}       \def\b{\beta}      \def\d{\delta}

                    \def\s{\sigma}
           
        \def\o{\omega}   
\def\G{\Gamma}

\def\AA{\buildrel_{\circ}\over{\mathrm{A}}}

\def\=={\equiv}

\def\qed{\raise1pt\hbox{\vrule height5pt width5pt depth0pt}}

\def\cG0{{\cal G}_0} 
\def\cG{{\cal G}}

 \def\=={\equiv}

 \def\ep0{\epsilon_{p}} \def\ed0{\epsilon_{f}}

\def\be{\begin{equation}}
\def\ee{\end{equation}}

\def\cc{c^{\dagger}}
\def\ca{c^{\phantom{\dagger}}}

\def\bc{b^{\dagger}}
\def\ba{b^{\phantom{\dagger}}}

\newcommand{\ket}[1]{|{#1}\rangle}

\newcommand{\quave}[1]{\langle{#1}\rangle}

\def\tns{Ta$_2$NiSe$_5$~}
\def\tnscomma{Ta$_2$NiSe$_5$}
\def\tplus{{\rm Ta}_+}
\def\tmins{{\rm Ta}_-}
\def\tpm{{\rm Ta}_\pm}

\def\nik{\rm Ni}
\def\PsiEI{\Psi_{\rm EI}}
\def\PhiTNTN{\Phi_{\lozenge}}

\usepackage{cancel}

\usepackage{fancyhdr}
\begin{document}

\author{Giacomo Mazza \\
{\it Dipartimento di Fisica dell'Universit\`a di Pisa, Largo Bruno Pontecorvo 3, I-56127 Pisa,~Italy}}

\title{
Magnetic field control of the excitonic transition in \tns
}	

\begin{abstract}
The formation of excitonic insulator phases in quantum materials 
is often masked by structural distortions caused by the coupling 
between electronic and phononic order parameters.
Here, for the candidate material \tnscomma, we show that sufficiently 
high magnetic fields can stabilize   
an excitonic insulating phase 
in the absence of a
 lattice distortion. 
By considering the interplay between the excitonic and structural 
instabilities, we predict a magnetic field-induced transition from 
the low-temperature, structurally distorted, semiconducting phase 
to an undistorted excitonic insulator phase with ground-state loop currents. 
Before the transition, the magnetic field-softening of the phonon mode 
associated with the structural distortion reveals the existence of 
a latent excitonic phase under the structurally distorted one. 
These results highlight an unbiased route towards the disentanglement 
of the coupled excitonic-structural transition in \tnscomma, and uncover 
a general mechanism for magnetic field control of competing 
phases in quantum materials.
\end{abstract}

\maketitle

\section{Introduction}
Tunable metal-insulator transitions (MITs) 
have enormous potential for technological 
applications of quantum materials~\cite{tokura_quantum_materials_2017,
resistive_switch_nature_reviews2020,del_valle_neuromorphic_2018}. 
Often, MITs suggest the formation of collective 
states of matter driven by strong electronic 
correlations, such as the Mott insulator state~\cite{imada_MITs_RMP1998,mott_original_1949}. 
However, in many correlated materials, MITs occur 
simultaneously with structural distortions that 
mask the true microscopic origin of the MIT 
and, in turn, deeply alter the nature 
of the correlation-driven insulating states.

Excitonic insulators (EIs) provide a paradigmatic 
example of MITs coupled to structural distortions. 
The EI state is the particle-hole analog 
of the superconducting paired state~\cite{kohn_excitonic_ins,kozlov_divalent_crystal,halperin_rice_rmp,
keldysh_collective_properties_excitons} occurring, for example, 
in bilayer systems with spatially separated particles and holes~\cite{spielman_double_quantum_hall_ferromagnet_prl2000,
kellogg_double_electron_gas_prl2004,eisenstein_nature2004,eisenstein_review_2014,
excitonic_insulator_atomic_double_layers_Nat2021,quantum_oscillation_dipolar_excitons_NatMat2025}.
However, despite the many hints~\cite{exp_ei_Ta2NiSe5,kogar_TiSe2,varsano_NatNano_2020,varsano_pnas_2021,
ei_monolayer_wte2_NatPhys_2022,
EI_wte2_NatPhys2022,
ei_black_phosphorous_2025,hossainTPT_nat_phys_2025}, the 
very existence of a bulk material hosting an EI phase 
remains a puzzle.
In fact, the physical properties of the EI state are not 
unambiguously defined. 
Rather, the EI instability represents a general proxy for 
breaking some of the internal symmetries of a solid,
so that there may exist many different EI states 
associated with different symmetry-breaking 
channels~\cite{giacomo_hiddenTRSB2023,
breakdown_topological_TRSB_papaj_2023,amaricci_QSHI2023,
spectroscopic_signatures_papaj_prb2024,Van-Nham_2025,nasu_magnetic_field_effect}. 
In particular, the EI instability often involves the 
breaking of discrete symmetries of the crystal lattice, 
so that concurring structural distortions represent 
an inevitable obstacle in its identification.

\begin{figure}[b]
\includegraphics[width=\columnwidth]{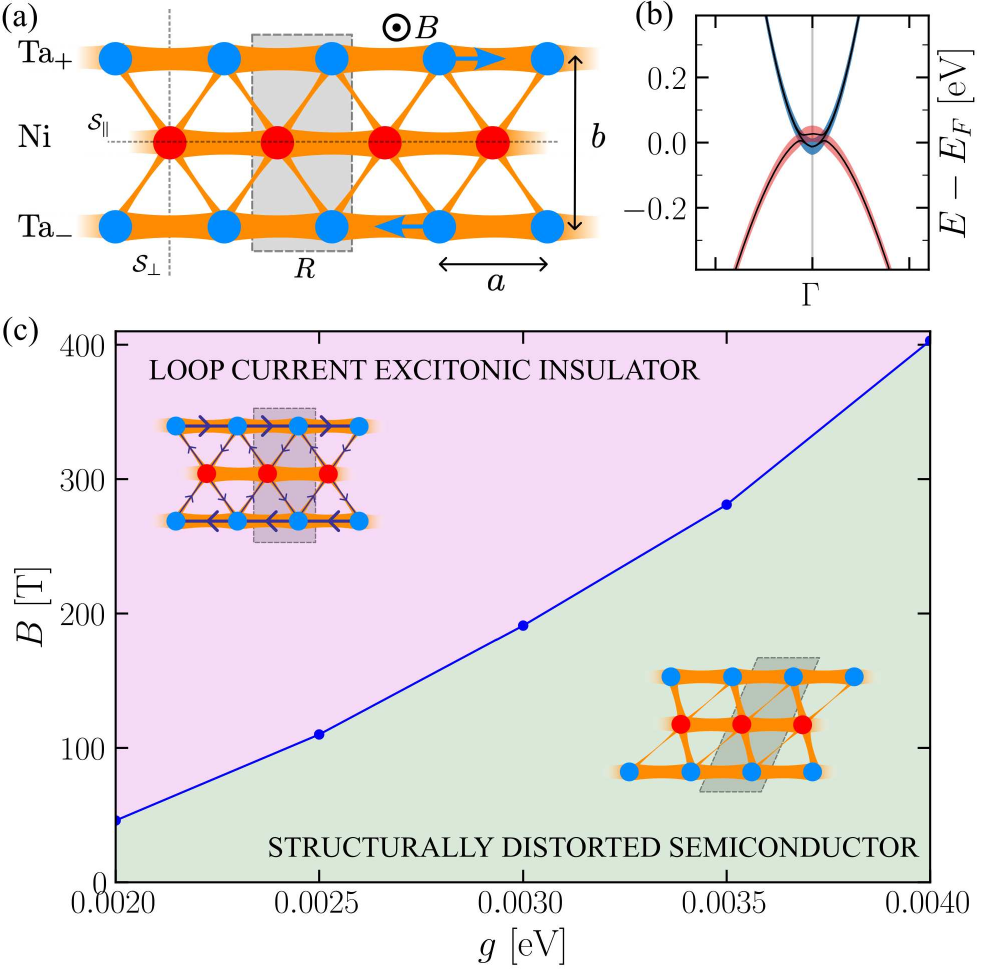}
\caption{(a) Ta-Ni-Ta chain structure in the high-temperature 
orthorombic phase.
The orange links schematically represent the charge distribution 
on the bonds. Shaded area highlights the unit cell.  
The blue arrows indicates the Ta-shear mode distortion.
$a \simeq 3.5~{\rm \AA}$ and $b \simeq 3.9~{\rm \AA}$ 
are the lattice parameters of the chain. 
(b) Low-energy band structure for the 
high-temperature phase tight-binding model.
Blue/red fat lines indicate the Ta/Ni characters of the bands.
(c) Zero temperature phase diagram as a function of the perpendicular 
field $B$, and of the electron-phonon coupling $g$.}
\label{fig:fig0}
\end{figure}

\tns is a candidate material in which the MIT has been 
associated to the formation of an EI state below a critical temperature, 
as suggested by the gap opening with characteristic flattening 
of the valence band around the $\G$-point~\cite{disalvo_Ta2NiSe5,seki_TNS,exp_ei_Ta2NiSe5}.
However, the EI transition breaks the same crystal symmetries 
of the orthorombic-to-monoclinic lattice distortion~\cite{giacomo_TNS,kaneko_ortho_to_mono,watson_tns2020} 
that occurs concomitantly with the MIT, stimulating  an intense debate about the 
excitonic or structural origin of the low-temperature insulating phase~\cite{journal_club_chicken_egg}.
So far, the various attempts to isolate the leading excitonic 
or structural character of the MIT, focusing, for example, on the analysis of collective excitations~\cite{kaiser_TNS,golez_nonlinear_spectroscopy_prl2020,
Kim2021_raman_TNS,Ye2021_raman_TNS,subedi_TNS,lukas_TNS,volkov_npjQM_2021,
golez_unveiling_2022,chen20252dcoherentspectroscopysignatures,
chatterjee_collective_modes_2025,Xuan_collective_modes_EI_2025,
katsumi_disentangling_PRL2023,baldini_TNS_pnas2023}, have not provided a definite answer. Recent investigations using gate doping~\cite{wei_gate_tuningTNS}, 
uniaxial pressure~\cite{uniaxial_pressure_control},
elastocaloric measurements~\cite{rosenberg2025elastocaloricsignatureexcitonicinstability}, 
and scanning tunneling microscopy~\cite{bae2025microscopicevidencedominantexcitonic} reached different conclusions. 

In this Article, we propose 
a solution to the conundrum based on the stabilization of 
a latent EI phase immune to the lattice distortion. 
We show that, if the MIT in \tns underlies an excitonic 
mechanism, the low-temperature structurally distorted phase 
can be transformed back into an undistorted EI by the application 
of a perpendicular magnetic field.

We first show that excitonic instability in \tns can 
either result in a time-reversal invariant charge distorted 
excitonic phase or a charge symmetric time-reversal symmetry 
broken phase with ground state loop currents.
Without the applied magnetic field, the electron-phonon 
coupling $g$ selects the charge distorted phase and drives 
the monoclinic distortion, thus transforming 
the excitonic state into a simple, structurally 
distorted, semiconductor. Application of the magnetic 
field suppresses the monoclinic distortion.
For a field larger than a critical one, $B>B_c$, 
the monoclinic distortion vanishes, 
signaling the restoration of the orthorombic structure and 
the stabilization of the excitonic insulating phase 
characterized by closed current loops within the 
unit cell, see Fig.~\ref{fig:fig0}. 
Such a loop-current phase  
corresponds to the purely excitonic insulating phase
which, in the absence of the applied field, is 
destabilized by the electron-phonon coupling. 
In the whole structurally distorted region of the 
phase diagram, the latent loop current EI phase manifests 
through a field-induced suppression of the monoclinic distortion 
and the magnetic field softening of the associated phonon mode. 

In Sec.~\ref{sec:EI_woutBfield}, we discuss the excitonic instability in the 
absence of applied field, and the coupling to the lattice leading 
the structural distortion. In Sec.~\ref{sec:EI_wBfield}, we show the restoration of symmetry upon application of a perpendicular magnetic field, 
and discuss possible experimental probe via magnetic-field induced 
phonon softening.

\section{Excitonic phases and structural distortion.}
\label{sec:EI_woutBfield}
\tns has a layered structure with each layer characterized by weakly 
coupled one-dimensional structures of alternating Ta-Ni-Ta chains, 
see Fig.~\ref{fig:fig0}(a).
In the high-temperature phase, the unit cell is orthorombic 
and the system is metallic.  
The transition towards the low-temperature insulating phase 
is well described by a coupled excitonic-structural 
transition within a minimal model of interacting electrons 
in the Ta and Ni chains coupled to the Ta-shear 
mode~\cite{kaneko_ortho_to_mono,giacomo_TNS,sugimoto_strong_couplingTNS}.
The full Hamiltonian, $H = H_{el} + H_{ph} + H_{el-ph}$, 
includes the purely electronic term  $H_{el} = H_0 + H_{int}$,  
containing the tight-binding, $H_0$, and density-density interaction, $H_{int}$, terms, 
the bare phonon Hamiltonian, $H_{ph}$, and the 
electron-phonon interaction, $H_{el-ph}$.
We restrict to a single Ta-Ni-Ta chain since hopping 
in the direction transverse to the chains is negligibly small 
as compared to the electronic dispersion along the 
longitudinal direction~\cite{giacomo_TNS,lukas_TNS}.
In the orthorombic phase, the tight-binding Hamiltonian reads
\begin{equation}
H_0 = \sum_{\s} \sum_{\a \b }^{\left\lbrace \tpm,\nik\right\rbrace} 
\sum_{R R'} t_{R,R'}^{\a \b} 
\cc_{R \a \s} \ca_{R' \b \s},
\label{eq:Htightbinding}
\end{equation}
where $\cc_{R \a \s}/\ca_{R \a \s }$, 
are creation/annihilation operators for electrons with spin 
$\s$ in the Wannier orbitals localized on the atom $\a $ 
in the unit cell $R$ which contains an upper $(\tplus)$ and 
a lower $(\tmins)$ Ta-atom, and a central Ni-atom.
In the following, we truncate the hoppings at the nearest-neighboring 
atoms, and fix $t_{R,R+a}^{\tplus,\tplus} = t_{R,R+a}^{\tmins,\tmins} = -0.72~{\rm eV}$,
$t_{R,R+a}^{\nik,\nik} = 0.3~{\rm eV}$, 
$t_{R,R}^{\tplus,\tplus} = t_{R,R}^{\tmins,\tmins} = 2.0~{\rm eV}$, 
$t_{R,R}^{\nik,\nik} = 0~{\rm eV}$, 
and $t_{R,R}^{{\rm T}_{\pm},\nik} = -t_{R,R-a}^{{\rm T}_{\pm},\nik}  = 35~{\rm meV}$,
obtaining a low-energy band structure characterized by weakly 
overlapping conduction and valence bands of predominant 
Ta- and Ni-characters respectively, see Fig.~\ref{fig:fig0}(b). 
The density-density interaction reads 
\begin{equation}
H_{int} = \frac{1}{2} \sum_{\s \s'} \sum_{\a \b} \sum_{R R'} 
U_{\a \b}^{\s \s'} (R-R') \cc_{R \a \s} \ca_{R \a \s} \cc_{R' \b \s'} \ca_{R' \b \s'},
\label{eq:Hint}
\end{equation}
where $U_{\a \b}^{\s \s'} (R-R')$ are  
long-range interaction potentials parametrized by local 
intra- and inter-chain interaction strengths 
$U_{\a} \equiv U_{\a \a}^{\s -\s}(R=R')$, $V \equiv U_{\tpm,\nik}^{\s \s'} (R = R')$,
and we neglected inter-chain $\tplus-\tmins$ interactions.

The bare phonon Hamiltonian reads $H_{ph} = \hbar \o_0 
\sum_{R} \sum_{\a = \tpm} \bc_{R \a} \ba_{R \a},$
where  $\bc_{\tpm R}/\ba_{\tpm R}$ are bosonic 
creation/annihilation operators
describing the vibration of the $\tpm$-atoms along the chains.
Finally, the electron-phonon Hamiltonian $H_{el-ph} = \sum_{\a = \tpm} H_{el-ph}^{\a}$ 
includes the linear coupling between $\tpm$ displacement and the 
$\tpm$-$\nik$ nearest neighbourg hybridization as 
\begin{equation}
H^{\a}_{el-ph} = g\sum_{R \s} 
X_{R \a} \left( \cc_{R \nik \s}  \ca_{R \a \s} + \cc_{R \nik \s}  \ca_{R-a \a \s} + {\rm h.c.} \right),
\label{eq:Heph_alpha}
\end{equation}
with $X_{R \pm} \equiv \bc_{R \pm}+\ba_{R \pm}$
and $g$ is a dimensional electron-phonon coupling.

We solve the electron-phonon model using the mean-field 
decoupling $\ket{\Psi} = \ket{\Psi_{el}} \ket{\Psi_{ph}}$, where 
$\ket{\Psi_{el}}$ and $\ket{\Psi_{ph}}$ are self-consistently 
determined  as the ground states of effective purely electronic, $\tilde{H}_{el}$, 
and phononic, $\tilde{H}_{ph}$, Hamiltonians obtained upon 
averaging the full Hamiltonian over the phononic and 
electronic wavefunctions, respectively. 
The electronic part is treated using a standard Hartree-Fock 
(HF) decoupling of the density-density interaction 
in the particle-hole channel 
upon fixing the average number of particles to 
two electrons per unit cell 
$n_{R} = \sum_{\a \s} \quave{\cc_{R \a \s} \ca_{R \a \s}} = 2$, 
see~\cite{suppl} for details.
In the following, we choose model parameters 
allowing for an excitonic instability.
We fix the intra-chain interaction parameters 
$U_{\nik} = 1.5~{\rm eV}$, $U_{\tpm} = 0.5~{\rm eV}$ and 
the bare phonon energy $\hbar \o_0 = 10~{\rm meV}$,
and vary the inter-chain interaction parameter $V$ and 
the electron-phonon coupling $g$.
Different choices of the model parameters do not 
alter the qualitative picture described in this 
work as long as the model parameters 
allow for an excitonic phase.

\begin{figure}
\includegraphics[width=\columnwidth]{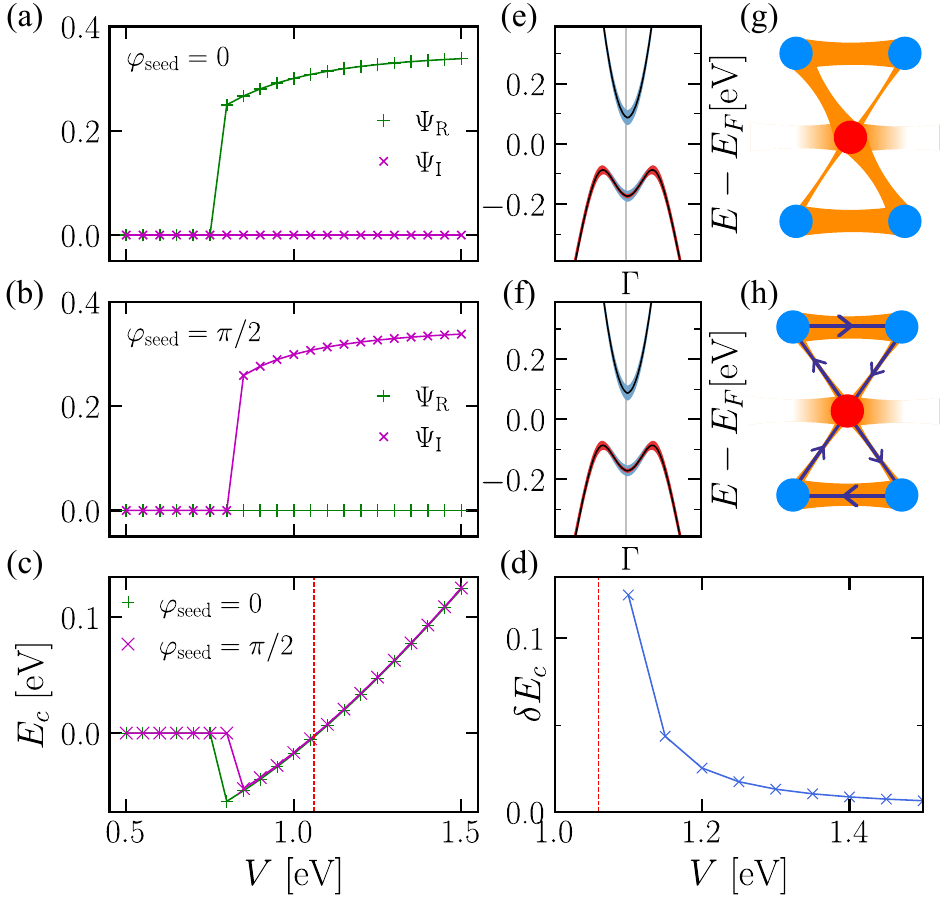}
\caption{(a)-(b) Real $\Psi_{\rm R}$ and imaginary $\Psi_{\rm I}$ parts
of the excitonic order parameter obtained using seeds $\Psi_{\rm seed} = 0.2e^{i \varphi_{\rm seed}}$
with $\varphi_{\rm seed}=0$ (a) and $\varphi_{\rm seed}=\pi/2$ (b). 
(c) Condensation energies for the two phases.
The vertical dashed lines highlights the
first order transition between the symmetric and symmetry broken phases. 
(d)  Relative energy difference $\d E_c = 1 - E_c(\varphi=0)/E_c(\varphi=\pi/2)$. (e)-(f) 
Low-energy bands in the $\varphi=0$ (e) and $\varphi=\pi/2$ (f) 
cases for $V=1.1~{\rm eV}$.
(g)-(h) Sketches of the charge density and the bond current distributions 
in the $\varphi=0$ (g) and $\varphi=\pi/2$ (h) phases.}
\label{fig:fig1}
\end{figure}

Let us first set $g=0$, and consider the purely 
excitonic instability. The EI instability in \tns 
is related to the breaking of the reflection symmetry 
with respect to the chain transverse plane ${\cal S}_{\perp}$ 
which constraints the hybridization between Ta- and Ni-Wannier 
orbitals to vanish at the $\G-$point. 
This implies the existence of a two-components complex 
order parameter 
\begin{equation}
\Psi_{\rm EI} = (\Psi_{+},\Psi_{-}),~{\rm with}~\Psi_{\pm } 
\equiv 
\sum_\s \quave{\cc_{k=0 \tpm \s} \ca_{k=0 \nik \s}},
\label{eq:psi_EI}
\end{equation}
and $\ca_{k \a \s} \equiv \frac{1}{N} \sum_{R} e^{i kR} \ca_{R \a \s}$, 
which is zero in the orthorombic phase, and  
becomes $\PsiEI \neq 0$ due to a spontaneous 
conduction/valence band hybridization triggered 
by the excitonic instability.
We assume that the instability preserves the 
symmetry of the low-temperature monoclinic 
phase, $\Psi_+=-\Psi_- = \Psi$.
This choice reduces the order parameter 
to a single complex number $\Psi = |\Psi| e^{i \varphi}.$

We solve the self-consistent HF problem at zero-temperature 
by seeding a non-zero excitonic order parameter in the complex plane with different phases, and find that, for $V \gtrsim 0.8~{\rm eV}$, 
symmetry breaking can either occur with a 
purely real, $\varphi = 0$, or purely imaginary, $\varphi=\pi/2$, 
order parameters, see Fig.~\ref{fig:fig1}(a)-(b).
In panel (c), we show the condensation energy 
$E_{c} (\Psi) \equiv E(\Psi=0) - E(\Psi) $ 
as a function of $V$.
For both order parameters the condensation 
energy becomes $E_c > 0$ for $V \gtrsim 1.05~{\rm eV}$
highlighting a first-order transition towards 
the symmetry broken excitonic phase.

Notably, the two excitonic phases are essentially 
indistinguishable from a band structure point of view.
In both cases, the symmetry breaking results 
in the opening of a slightly indirect gap which, 
for $V=1.1~{\rm eV}$, is $\sim 0.15~{\rm eV}$, see Fig.~\ref{fig:fig0}. 
However, by considering a symmetry analysis 
of charge and current densities, one can show 
that the two EI phases correspond to two strikingly 
different physical states.
In fact, the excitonic order parameter controls the 
hybridization between neighbouring Wannier orbitals 
and, since the latter are real, the real and imaginary 
parts of the hybridization affect, respectively, 
the charge and current densities on the bonds between 
two Wannier centers, see~\cite{suppl} for details.
Specifically, the $\varphi=0$ state corresponds to 
the usual charge distorted configuration 
of \tns \cite{watson_tns2020,giacomo_TNS} 
characterized by a charge density asymmetry on the 
neighboring $\tpm$-Ni bonds, see Fig.~\ref{fig:fig1}(g). 
On the contrary, for $\varphi=\pi/2$ the charge 
density preserves the symmetries of the orthorombic 
phase whereas the symmetry breaking results 
in the formation of a loop current pattern 
on the $\tpm$-$\nik$ and $\tpm$-$\tpm$ bonds, see Fig.~\ref{fig:fig1}(h).
In the following, we dub the 
excitonic phases characterized by 
a purely real and a purely imaginary order 
parameter as the charge distorted (CD) 
and the loop current (LC) excitonic 
phases, respectively. 

\begin{figure}
\includegraphics[width=\columnwidth]{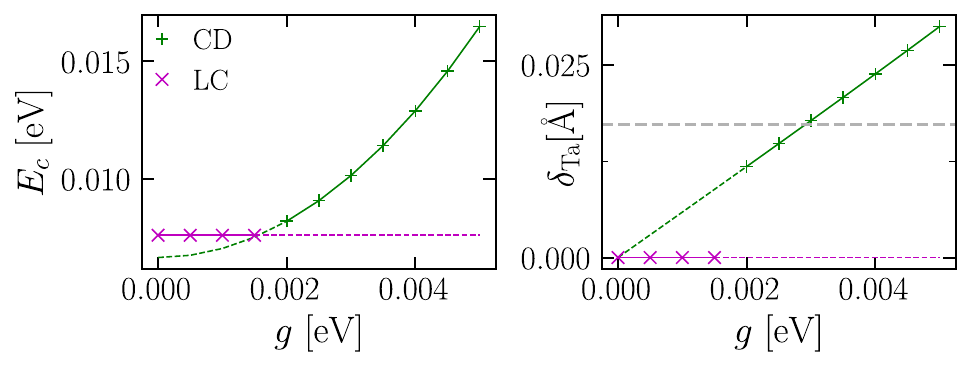}
\caption{(a) Condensation energy of the CD and LC phases as a 
function of the dimensional electron-phonon couplingfor $V=1.1~{\rm eV}$.
For the CD and LC phases, the dashed lines indicate 
that, for the corresponding values of $g$, 
the stationary solution is unstable. 
(b) Displacement of the Ta-atom with respect to the 
equilibrium position in the orthorombic phase.
The gray dashed lines indicates the 
displacement corresponding to 
a monoclinic angle of $0.5^{\circ}$.
}
\label{fig:fig2}
\end{figure}

In the presence of two possible symmetry breaking 
channels, we identify the stable phase by 
comparing the energy of the two phases and by
carrying out a stability analysis of the stationary solutions, 
see ~\cite{suppl} for details.
Interestingly, we find that the LC always 
corresponds to the lower energy stationary solution. 
Moreover, the LC solution corresponds to a stable 
energy minimum, whereas the CD solution is an unstable saddle point.
We therefore identify the LC as the only stable 
symmetry broken phase in the purely electronic model.

The above observation has some relevant consequences. 
First of all, we notice that the stable LC phase 
has been so far overlooked in similar microscopic models for \tnscomma.
For completeness, we checked the robustness of this 
observation by showing that the LC phase is the 
stable one also in the double chains model of \tns with 
short range interaction originally introduced 
in Ref.~\cite{giacomo_TNS}, see \cite{suppl}.
Most importantly, the stability of the LC phase 
shows that the excitonic instability alone 
can hardly explain the formation of the 
low-temperature symmetry broken phase in \tnscomma. 
In fact, even if assuming that the 
energetic competition between 
the LC and CD phases may actually depend on
some details of the microscopic model, 
the fact that the nearly degenerate LC and CD
solutions corresponds to extremely different 
physical states strongly suggests that the 
realization of the actual symmetry-broken 
phase unavoidably depends on the coupling 
with other degrees of freedom. 
In this scenario, the Ta-chains shear mode 
distortion is not a mere spectator 
of the excitonic instability, but it 
becomes instrumental to select the CD phase
and overrule the competition between the 
different excitonic states.  

In Fig.~\ref{fig:fig2}(a), we show the condensation
energy of the CD and LC phases for $V=1.1~{\rm eV}$, 
and as a function of the electron-phonon coupling $g$.
As expected, the coupling with phonons enhances 
the condensation energy of the CD phase, and 
induces a static monoclinic distortion 
$ \quave{X_{R+}} = -\quave{X_{R-}} \neq 0$ 
that grows linearly with $g$, Fig.~\ref{fig:fig2}(b).
On the contrary, the energy of the LC phase does 
not depend on $g$ and the mode remains undistorted. 
This follows from the absence of a linear coupling 
between the LC order parameter and the shear mode 
distortion which are, respectively, odd and even 
under time-reversal symmetry.
For $g \gtrsim 1.5~{\rm meV}$, the condensation 
energy of the CD state becomes larger than 
the one of the LC phase. By the stability analysis, 
we confirm that the energy crossing coincides with 
the transformation of the CD saddle point for $g=0$ 
into an energy minimum for $g>g_c$,  and the simultaneous  
destabilization the LC phase which becomes a saddle point for $g>g_c$, 
see~\cite{suppl}.
By estimating the the Ta-atom 
displacement, as $\delta_{\rm Ta} = \ell_0 \quave{X_{R+}} $ 
with $\ell_0 = \sqrt{{\hbar}/{(2 M_{\rm Ta} \o_0)}} \approx 0.03~{\rm \AA}$, 
being $M_{\rm Ta}$ the mass of the Ta-ion, 
we see that the observed monoclinic angle 
$\sim 0.5^{\circ}$~\cite{disalvo_Ta2NiSe5} 
falls well within the region of the stable 
distorted CD phase for $g \sim 3~{\rm meV}$.
Eventually, we comment on the fact that the 
stabilization of the CD phase may be further 
enhanced by higher order symmetry-allowed 
coupling, such as $\sim X^2 \Psi_{\rm I}^2$,
which will results in a decrease 
of the condensation energy of the LC phase 
and a shift the transition to smaller 
values of $g$ without changing the overall picture.

\section{Field-induced stabilization of the undistorted EI phase.}
\label{sec:EI_wBfield}
Despite the fact that the electron-phonon coupling 
destabilizes the LC phase for $g \gtrsim 1.5~{\rm meV}$, 
the very existence of such a stable phase
for $g=0$ suggests that it may be possible to counterbalance 
the detrimental effects of the coupling to the lattice 
by using a perpendicular magnetic field 
which transforms in the same way as 
the LC order parameter under both time-reversal and 
reflection symmetries.
We include a constant magnetic field, $B$, via Peierls phases in the hoppings
\begin{equation}
t_{R,R'}^{\a \b} \to  t_{R,R'}^{\a \b} e^{i \phi_{R, R'}^{\a,\b}/\Phi_0} ,
\label{eq:peierls}
\end{equation}
with $\Phi_0 \equiv \frac{\hbar}{e}$.
The Peierls phases are defined in terms of the flux 
through the $\tplus$-Ni-$\tmins$-Ni diamond-like plaquette 
$\Phi_{\lozenge} = B A_{\lozenge}$, with $A_{\lozenge} = ab \simeq 0.14~{\rm nm}^2$.
We fix the relative phases by imposing  
$\PhiTNTN  = 2 \Phi_{\bigtriangledown} = 2 \Phi_{\bigtriangleup} $,
$\Phi_{\bigtriangledown/\bigtriangleup}$ being the fluxes threading the 
upper $(\tplus$-$\nik$-$\tplus)$/lower $(\tmins$-$\nik$-$\tmins)$ triangular plaquettes, 
and obtain
$\phi_{R,R}^{\tplus \nik} = \frac{\PhiTNTN}{4}$,  $\phi_{R-a,R}^{\tplus \nik} = -\frac{\PhiTNTN}{4}$, 
$\phi_{R+a,R}^{\tplus \tplus} = -\phi_{R+a,R}^{\tmins \tmins} = \PhiTNTN$, 
and $\phi_{R+a,R}^{\nik \nik} = 0.$
In the following, we will consider ultra high fields 
up to few hundreds Tesla. Despite that, the flux 
through the plaquette is always $\PhiTNTN \ll \Phi_0$
so that interaction terms beyond the Peierls phases 
can be safely neglected~\cite{peierls_original,
luttinger_peierls_phases1951,
electromagnetic_coupling_2020}.
Moreover, we estimate the Zeeman 
splitting effects by considering a g-factor 
$g_s = 2$. 
For $B=300~{\rm T}$, the Zeeman splitting 
is of the order of $0.02~{\rm eV}$, 
whereas the electronic gap is of the order of 
$\gtrsim 0.15~{\rm eV}$, so that, upon inclusion 
of the Zeeman splitting, we obtain a zero 
spin magnetic-moment for all the considered values of $B$.
Eventually, we neglect spin-orbit 
couplings effect as these are known to be weak 
and negligible along the direction of the chains, see Ref.~\cite{lukas_TNS}.

\begin{figure}
\includegraphics[width=\columnwidth]{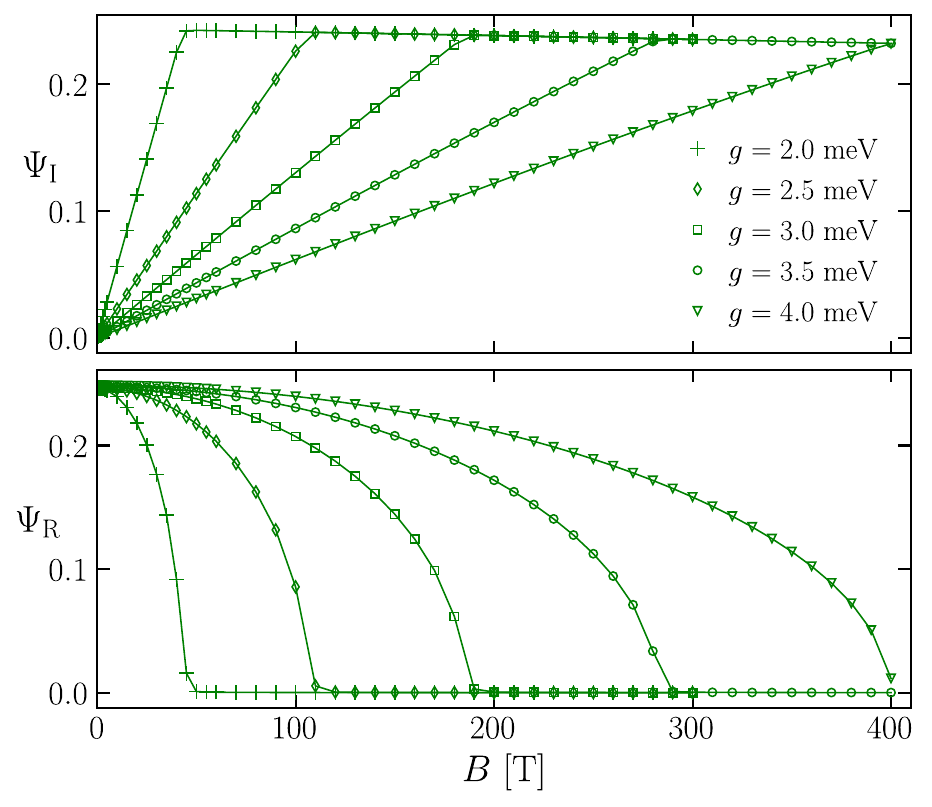}
\caption{Imaginary (top) and real (bottom) components 
of the excitonic order parameter as a function of the applied
magnetic field $B$, and different values 
of the electron phonon coupling $g$. $V=1.1~{\rm eV}$.}
\label{fig:fig3}
\end{figure}

The applied flux couples linearly with the imaginary 
part of the order parameter as can be readily checked 
by expanding Peierls factors for the $\tpm-\nik$ 
nearest-neighbor hoppings at linear order 
\begin{equation}
\d H_{\a} \simeq 
\sum_{R \s} 
i \frac{\phi_{R,R}^{\a \nik}}{\Phi_0}  t^{\a,\nik}_{R,R}
\left( \cc_{\nik R \s} \ca_{\a R \s} + \cc_{\nik R \s} \ca_{\a R-a \s} - 
{\rm h.c.}   \right),
\end{equation}
with $\a=\tpm$, and 
$\phi^{\tpm \nik}_{R,R} = \pm \frac{\PhiTNTN}{4}$ previously 
defined.
Due to the linear coupling, the applied flux 
acts as a symmetry breaking field for the imaginary 
component of the order parameter associated with 
the LC phase. In Fig.~\ref{fig:fig3}, 
we show the imaginary and real components
of the order parameter as a function 
of $B$, starting from the structurally 
distorted phase and different values of $g$.
The imaginary component increases linearly as 
a function $B$, whereas the real component gets suppressed. 
This can be easily understood with the help of 
a Landau functional for the real and 
imaginary components of the order parameter 
with lowest order symmetry-allowed couplings,
$F [\Psi_{\rm R},\Psi_{\rm I}] 
= a_{\rm R} \Psi_{\rm R}^2 + b_{\rm R} \Psi_{\rm R}^4 + 
	a_{\rm I} \Psi_{\rm I}^2 + b_{\rm I} \Psi_{\rm I}^4 + c \Psi_{\rm I}^2 
	\Psi_{\rm R}^2 + \ldots$, 
	where, $a_{\rm R} <0$ and $a_{\rm I} > 0$ being in the CD phase, 
	and $b_{\rm R/I}>0$, and $c>0$ for stability.
	Due to the non linear coupling $\Psi_{\rm I}^2 \Psi_{\rm R}^2$,  
	the field-induced growth of $\Psi_{\rm I}$
	creates an effective mass term 
	that is responsible of the quadratic 
	suppression of real component $\Psi_{\rm R}$.  
	The simultaneous field-induced growth/suppression 
	of the order parameter components associated, respectively, 
	with the LC and CD phases continues up to a $g-$dependent 
	critical value $B_c$ at which the real component  
	continuously goes to zero and the imaginary 
	one collapses to a $g-$independent curve.
	By the stability analysis, we checked that stable energy 
	minimum continuously shifts from the purely real order 
	parameter to the purely imaginary as the applied 
	field approaches the critical value $B_c$~\cite{suppl}.
		
\begin{figure}
\includegraphics[width=\columnwidth]{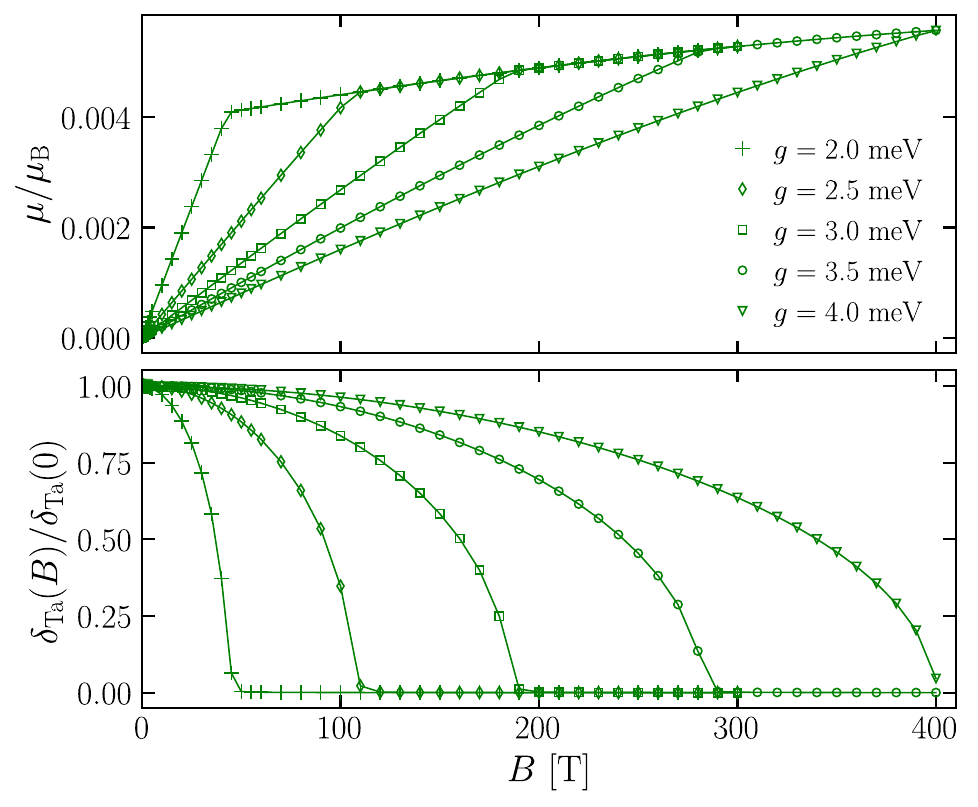}
\caption{(a) Orbital magnetic moment 
as a function of the applied magnetic field $B$, 
and for different values of the electron phonon coupling  $g$.
(b) Static displacement of the Ta-atoms 
as a function of the applied fields and the 
same values of $g$ displayed in panel (a). $V=1.1~{\rm eV}$.}
\label{fig:fig4}
\end{figure}

	It is immediate to check that the 
	field-induced stabilization of a the imaginary 
	component of the order parameter and the 
	concomitant suppression 
	of the real one, coincides with the stabilization of the 
	undistorted LC excitonic phase.
	In Fig.~\ref{fig:fig4}, we plot
	the magnetic field evolution of the 
	two observables directly associated to the 
	LC and CD components of the order parameter, respectively.
	In panel (a), we plot the orbital magnetic moment, 
	defined as $\vec{\mu} = \int d\vec{r}~\vec{r} \times \vec{j}(\vec{r})$,
	where $\vec{r}$ is the vector originating from the 
	Ni-atom of the $R=0$ cell, and $\vec{j}(\vec{r})$ the 
	current density defined on the links~\cite{suppl}. Panel (b) 
	displays the shear mode distortion.
	The magnetic moment linearly increases with a 
	a paramagnetic susceptibility decreasing as a function of $g$, and 
	eventually merges into a single curve for $B>B_c$. 
	The mode distortion closely follows the evolution 
	of the real component of the excitonic order parameter, 
	and vanishes for $B>B_c$ showing  that a sufficiently high 
	magnetic-field can revert the orthorombic-to-monoclinic 
	distortion and stabilize the undistorted excitonic 
	LC phase characterized by a finite orbital magnetization.
					
\begin{figure}
\includegraphics[width=\columnwidth]{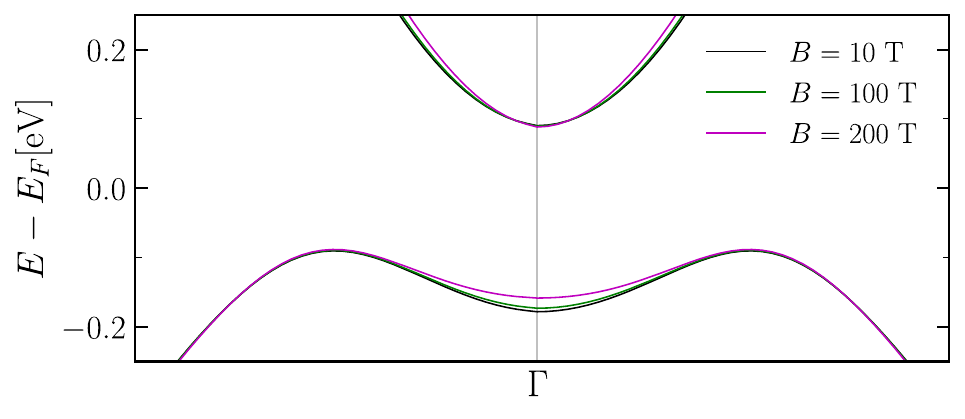}
\caption{Electronic bands across the magnetic-field induced transition
for $g=3~{\rm meV}$ and $V=1.1~{\rm eV}$.}
\label{fig:fig5}
\end{figure}

	The fields required to stabilize the undistorted 
	LC phase depends on the coupling $g$ and can reach 
	a few hundreds of Tesla for $g \gtrsim 0.3~{\rm meV}$.
	Despite that, the flux through the plaquette remains 
	very small reaching, for example, $\PhiTNTN \sim 0.01 \Phi_0$
	for $B=300~{\rm T}$.
	As a result, the electronic spectrum evolves smoothly 
	across the transition, see~Fig.~\ref{fig:fig5}. 
	In particular, the electronic gap remains unchanged 
	as a direct consequence of the fact that, as already observed, 	
	the CD and LC phase have almost indistinguishable electronic spectra.

\subsection{Phonon softening}
The magnetic field induced suppression 
of the structural distortion represents a 
clear evidence of the excitonic nature of the 
low-temperature phase of \tnscomma.
In fact, this is only possible due to the existence of a 
latent LC excitonic phase which, in the absence of the applied 
field, gets destabilized by the coupling with the lattice.
Possible experimental verification includes 
the direct measurement of the lattice 
distortion suppression as a function of the applied fields.
However, this may be challenging due to 
the ultra high fields involved. 
For example, if we set $g=3~{\rm meV}$ 
by comparing the lattice distortion 
calculated in this model at zero-field with 
the observed one, we would obtain $B_c \simeq 190~{\rm T}.$
Moreover, this is made even more challenging due 
to the fact that the unperturbed monoclinic distortion 
is rather small, 
with a monoclinic angle $\sim 0.5^{\circ}$~\cite{disalvo_Ta2NiSe5}. 

\begin{figure}
\includegraphics[width=\columnwidth]{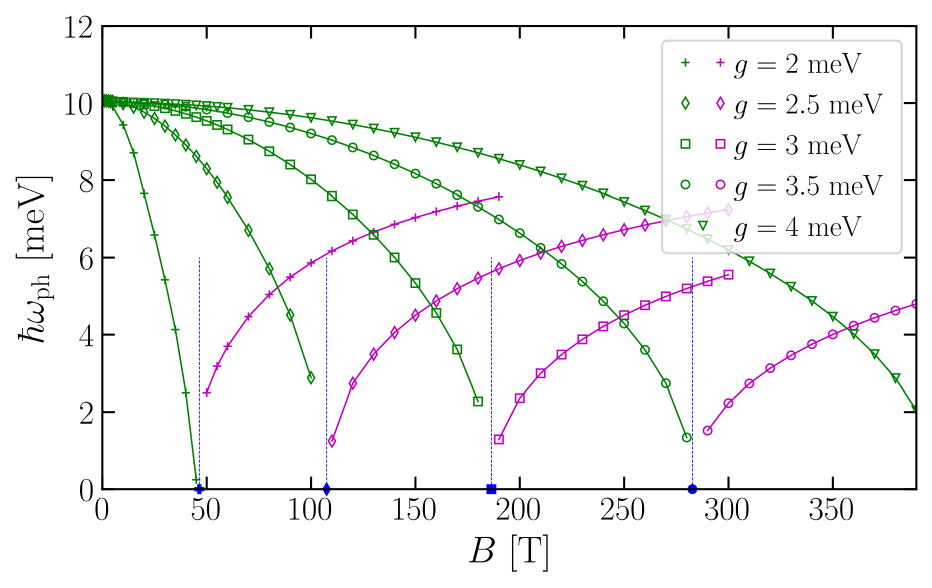}
\caption{Phonon frequency as a function of the magnetic field 
and different values of electron-phonon coupling $g$, and $V=1.1~{\rm eV}$.
For each $g$, the green and magenta symbols 
corresponds, respectively, to the structurally distorted 
semiconductor and LC excitonic insulator phase. 
The dots and the vertical dashed lines indicate 
the critical fields $B_c$.}
\label{fig:fig6}
\end{figure}

On the contrary, the magnetic field dependence of the 
shear mode frequency offers a powerful tool to probe the 
magnetic field-induced suppression of the distortion
at fields much smaller than the critical one.
The shear mode frequency is the collective 
excitation associated with the lattice distortion 
and it is expected to show critical softening 
at the continuous restoration of the orthorombic symmetry
for $B \to B_c$.
We compute the phonon spectral functions $A_{ph}(\o)$ 
using time-dependent HF and extract the phonon 
frequency as $\o_{ph} \equiv \int d\o A_{ph}(\o) \o / \int d\o A_{ph}(\o) $,
see~\cite{suppl} for details.
In Fig.~\ref{fig:fig6}, we show the phonon frequency 
dependence as a function of $B$ and increasing values of $g$. 
The phonon mode shows a standard critical softening 
behavior at a continuous quantum phase transition with
$\partial \o_{\rm ph}/\partial B < 0$, for $B < B_c$,
and $\o_{ph} \to 0$ as $B \to B_c$, 
followed by a phonon hardening, 
$\partial \o_{\rm ph}/\partial B > 0$, for $B> B_c$.
At the critical point, the mode frequency is characterized by 
a non-analytic cusp-like behavior for $|B-B_c| \to 0$.

The magnetic softening of the shear mode provides 
a clear cut criterion for the determination of the 
dominant mechanism underlying the MIT in \tnscomma. 
Specifically, if the dominant mechanisms is 
excitonic, we expect the mode to show a strong magnetic 
field softening related to the latent LC excitonic phase. 
Most importantly, since the phonon mode shows sizable 
softening for fields much smaller than the critical ones, 
we expect measurable signatures of magnetic mode 
softening already at much more reasonable magnetic 
fields.
For example, for $g=3~{\rm meV}$ we expect 
a $\sim 10\%$ softening, corresponding 
to $\Delta \o_{ph} \sim 1~{\rm meV}$, for $B \sim 65~{\rm T}$. 
On the contrary, if the excitonic mechanism is irrelevant, 
we expect the magnetic field dependence of the lattice 
response to be negligible up to very large applied fields.

\section{Conclusions}
\label{sec:conclusions}
We have shown the stabilization of a latent loop-current 
(LC) EI phase in \tns by application of perpendicular magnetic fields.
In the purely electronic model, the loop-current phase 
is the only stable excitonic phase, whereas the stability 
of the charge distorted phase is only enabled by the 
coupling with the lattice mode driving the monoclinic 
distortion.
The application of a perpendicular magnetic field 
suppresses the monoclinic distortion, and sufficiently
high fields restore the orthorombic structure and 
stabilize the LC excitonic phase in the absence of structural distortion.

We expect critical fields for stabilizing the undistorted LC phase 
to be of the order of a few hundred Tesla.
Tunability of MITs under such ultra-high fields has already 
been explored, for example, in vanadium oxides~\cite{magnetic_fild_V02_ncomm2020}.
Nonetheless, these results show that the magnetic field 
response of the lattice dynamics may be able to reveal the excitonic 
nature of the \tns low-temperature phase at fields much smaller 
than the critical ones.

The magnetic-field control of the excitonic transition
discussed in this work provides a new tool 
for disentangling excitonic and structural transition in \tnscomma.
In particularly, this is conceptually distinct to recent strain 
control experiments~\cite{uniaxial_pressure_control,
rosenberg2025elastocaloricsignatureexcitonicinstability},
in which the excitonic nature of \tns 
has been investigated by analyzing electronic properties 
following the direct suppression of the monoclinic distortion.
On the contrary, the results contained in this work 
show that the excitonic nature can be revealed by the magnetic 
response of the lattice distortion which can be only 
enabled by the existence of the, so far overlooked, 
loop-current excitonic phase. 

Besides being directly relevant in the case of \tnscomma, the mechanism for 
the stabilization of the LC phase due to interaction-driven spontaneous 
hybridization can be very general. In this respect, these results highlight 
the excitonic instability and its interplay with lattice distortion
as a promising route for the investigation of tunable competing phases 
in other quantum materials, such as loop-current phases 
in kagome metals~\cite{loop_current_kagome} or current 
whirlpools  in Chern insulators~\cite{cioni2026_persistent_currents}.

\begin{acknowledgments}
I thank M. R\"osner, A. Millis, A. Georges,
D. Gole\v{z}, A. Amaricci, M. Capone, and M. Polini 
for discussions. 
This work received funding from the MUR, Italian Minister 
of University and Research, through a 
``Rita Levi-Montalcini" fellowship. 
\end{acknowledgments}

\section*{DATA AVAILABILITY}
The data that support the findings of this article 
are openly available~\cite{mazza_2026_22232200}.


\bibliography{biblio_TNS}

\begin{thebibliography}{1}%
\makeatletter
\providecommand \@ifxundefined [1]{%
 \@ifx{#1\undefined}
}%
\providecommand \@ifnum [1]{%
 \ifnum #1\expandafter \@firstoftwo
 \else \expandafter \@secondoftwo
 \fi
}%
\providecommand \@ifx [1]{%
 \ifx #1\expandafter \@firstoftwo
 \else \expandafter \@secondoftwo
 \fi
}%
\providecommand \natexlab [1]{#1}%
\providecommand \enquote  [1]{``#1''}%
\providecommand \bibnamefont  [1]{#1}%
\providecommand \bibfnamefont [1]{#1}%
\providecommand \citenamefont [1]{#1}%
\providecommand \href@noop [0]{\@secondoftwo}%
\providecommand \href [0]{\begingroup \@sanitize@url \@href}%
\providecommand \@href[1]{\@@startlink{#1}\@@href}%
\providecommand \@@href[1]{\endgroup#1\@@endlink}%
\providecommand \@sanitize@url [0]{\catcode `\\12\catcode `\$12\catcode
  `\&12\catcode `\#12\catcode `\^12\catcode `\_12\catcode `\%12\relax}%
\providecommand \@@startlink[1]{}%
\providecommand \@@endlink[0]{}%
\providecommand \url  [0]{\begingroup\@sanitize@url \@url }%
\providecommand \@url [1]{\endgroup\@href {#1}{\urlprefix }}%
\providecommand \urlprefix  [0]{URL }%
\providecommand \Eprint [0]{\href }%
\providecommand \doibase [0]{http://dx.doi.org/}%
\providecommand \selectlanguage [0]{\@gobble}%
\providecommand \bibinfo  [0]{\@secondoftwo}%
\providecommand \bibfield  [0]{\@secondoftwo}%
\providecommand \translation [1]{[#1]}%
\providecommand \BibitemOpen [0]{}%
\providecommand \bibitemStop [0]{}%
\providecommand \bibitemNoStop [0]{.\EOS\space}%
\providecommand \EOS [0]{\spacefactor3000\relax}%
\providecommand \BibitemShut  [1]{\csname bibitem#1\endcsname}%
\let\auto@bib@innerbib\@empty
\bibitem [{\citenamefont {Mazza}\ \emph {et~al.}(2020)\citenamefont {Mazza},
  \citenamefont {R\"osner}, \citenamefont {Windg\"atter}, \citenamefont
  {Latini}, \citenamefont {H\"ubener}, \citenamefont {Millis}, \citenamefont
  {Rubio},\ and\ \citenamefont {Georges}}]{giacomo_TNS}%
  \BibitemOpen
  \bibfield  {author} {\bibinfo {author} {\bibfnamefont {G.}~\bibnamefont
  {Mazza}}, \bibinfo {author} {\bibfnamefont {M.}~\bibnamefont {R\"osner}},
  \bibinfo {author} {\bibfnamefont {L.}~\bibnamefont {Windg\"atter}}, \bibinfo
  {author} {\bibfnamefont {S.}~\bibnamefont {Latini}}, \bibinfo {author}
  {\bibfnamefont {H.}~\bibnamefont {H\"ubener}}, \bibinfo {author}
  {\bibfnamefont {A.~J.}\ \bibnamefont {Millis}}, \bibinfo {author}
  {\bibfnamefont {A.}~\bibnamefont {Rubio}}, \ and\ \bibinfo {author}
  {\bibfnamefont {A.}~\bibnamefont {Georges}},\ }\href {\doibase
  10.1103/PhysRevLett.124.197601} {\bibfield  {journal} {\bibinfo  {journal}
  {Phys. Rev. Lett.}\ }\textbf {\bibinfo {volume} {124}},\ \bibinfo {pages}
  {197601} (\bibinfo {year} {2020})}\BibitemShut {NoStop}%
\end{thebibliography}%


\begin{thebibliography}{61}%
\makeatletter
\providecommand \@ifxundefined [1]{%
 \@ifx{#1\undefined}
}%
\providecommand \@ifnum [1]{%
 \ifnum #1\expandafter \@firstoftwo
 \else \expandafter \@secondoftwo
 \fi
}%
\providecommand \@ifx [1]{%
 \ifx #1\expandafter \@firstoftwo
 \else \expandafter \@secondoftwo
 \fi
}%
\providecommand \natexlab [1]{#1}%
\providecommand \enquote  [1]{``#1''}%
\providecommand \bibnamefont  [1]{#1}%
\providecommand \bibfnamefont [1]{#1}%
\providecommand \citenamefont [1]{#1}%
\providecommand \href@noop [0]{\@secondoftwo}%
\providecommand \href [0]{\begingroup \@sanitize@url \@href}%
\providecommand \@href[1]{\@@startlink{#1}\@@href}%
\providecommand \@@href[1]{\endgroup#1\@@endlink}%
\providecommand \@sanitize@url [0]{\catcode `\\12\catcode `\$12\catcode
  `\&12\catcode `\#12\catcode `\^12\catcode `\_12\catcode `\%12\relax}%
\providecommand \@@startlink[1]{}%
\providecommand \@@endlink[0]{}%
\providecommand \url  [0]{\begingroup\@sanitize@url \@url }%
\providecommand \@url [1]{\endgroup\@href {#1}{\urlprefix }}%
\providecommand \urlprefix  [0]{URL }%
\providecommand \Eprint [0]{\href }%
\providecommand \doibase [0]{http://dx.doi.org/}%
\providecommand \selectlanguage [0]{\@gobble}%
\providecommand \bibinfo  [0]{\@secondoftwo}%
\providecommand \bibfield  [0]{\@secondoftwo}%
\providecommand \translation [1]{[#1]}%
\providecommand \BibitemOpen [0]{}%
\providecommand \bibitemStop [0]{}%
\providecommand \bibitemNoStop [0]{.\EOS\space}%
\providecommand \EOS [0]{\spacefactor3000\relax}%
\providecommand \BibitemShut  [1]{\csname bibitem#1\endcsname}%
\let\auto@bib@innerbib\@empty
\bibitem [{\citenamefont {Tokura}\ \emph {et~al.}(2017)\citenamefont {Tokura},
  \citenamefont {Kawasaki},\ and\ \citenamefont
  {Nagaosa}}]{tokura_quantum_materials_2017}%
  \BibitemOpen
  \bibfield  {author} {\bibinfo {author} {\bibfnamefont {Y.}~\bibnamefont
  {Tokura}}, \bibinfo {author} {\bibfnamefont {M.}~\bibnamefont {Kawasaki}}, \
  and\ \bibinfo {author} {\bibfnamefont {N.}~\bibnamefont {Nagaosa}},\ }\href
  {\doibase 10.1038/nphys4274} {\bibfield  {journal} {\bibinfo  {journal}
  {Nature Physics}\ }\textbf {\bibinfo {volume} {13}},\ \bibinfo {pages} {1056}
  (\bibinfo {year} {2017})}\BibitemShut {NoStop}%
\bibitem [{\citenamefont {Wang}\ \emph {et~al.}(2020)\citenamefont {Wang},
  \citenamefont {Wu}, \citenamefont {Burr}, \citenamefont {Hwang},
  \citenamefont {Wang}, \citenamefont {Xia},\ and\ \citenamefont
  {Yang}}]{resistive_switch_nature_reviews2020}%
  \BibitemOpen
  \bibfield  {author} {\bibinfo {author} {\bibfnamefont {Z.}~\bibnamefont
  {Wang}}, \bibinfo {author} {\bibfnamefont {H.}~\bibnamefont {Wu}}, \bibinfo
  {author} {\bibfnamefont {G.~W.}\ \bibnamefont {Burr}}, \bibinfo {author}
  {\bibfnamefont {C.~S.}\ \bibnamefont {Hwang}}, \bibinfo {author}
  {\bibfnamefont {K.~L.}\ \bibnamefont {Wang}}, \bibinfo {author}
  {\bibfnamefont {Q.}~\bibnamefont {Xia}}, \ and\ \bibinfo {author}
  {\bibfnamefont {J.~J.}\ \bibnamefont {Yang}},\ }\href {\doibase
  10.1038/s41578-019-0159-3} {\bibfield  {journal} {\bibinfo  {journal} {Nature
  Reviews Materials}\ }\textbf {\bibinfo {volume} {5}},\ \bibinfo {pages} {173}
  (\bibinfo {year} {2020})}\BibitemShut {NoStop}%
\bibitem [{\citenamefont {del Valle}\ \emph {et~al.}(2018)\citenamefont {del
  Valle}, \citenamefont {Ramirez}, \citenamefont {Rozenberg},\ and\
  \citenamefont {Schuller}}]{del_valle_neuromorphic_2018}%
  \BibitemOpen
  \bibfield  {author} {\bibinfo {author} {\bibfnamefont {J.}~\bibnamefont {del
  Valle}}, \bibinfo {author} {\bibfnamefont {J.~G.}\ \bibnamefont {Ramirez}},
  \bibinfo {author} {\bibfnamefont {M.~J.}\ \bibnamefont {Rozenberg}}, \ and\
  \bibinfo {author} {\bibfnamefont {I.~K.}\ \bibnamefont {Schuller}},\ }\href
  {\doibase 10.1063/1.5047800} {\bibfield  {journal} {\bibinfo  {journal}
  {Journal of Applied Physics}\ }\textbf {\bibinfo {volume} {124}},\ \bibinfo
  {pages} {211101} (\bibinfo {year} {2018})}\BibitemShut {NoStop}%
\bibitem [{\citenamefont {Imada}\ \emph {et~al.}(1998)\citenamefont {Imada},
  \citenamefont {Fujimori},\ and\ \citenamefont {Tokura}}]{imada_MITs_RMP1998}%
  \BibitemOpen
  \bibfield  {author} {\bibinfo {author} {\bibfnamefont {M.}~\bibnamefont
  {Imada}}, \bibinfo {author} {\bibfnamefont {A.}~\bibnamefont {Fujimori}}, \
  and\ \bibinfo {author} {\bibfnamefont {Y.}~\bibnamefont {Tokura}},\ }\href
  {\doibase 10.1103/RevModPhys.70.1039} {\bibfield  {journal} {\bibinfo
  {journal} {Rev. Mod. Phys.}\ }\textbf {\bibinfo {volume} {70}},\ \bibinfo
  {pages} {1039} (\bibinfo {year} {1998})}\BibitemShut {NoStop}%
\bibitem [{\citenamefont {Mott}(1949)}]{mott_original_1949}%
  \BibitemOpen
  \bibfield  {author} {\bibinfo {author} {\bibfnamefont {N.~F.}\ \bibnamefont
  {Mott}},\ }\href {\doibase 10.1088/0370-1298/62/7/303} {\bibfield  {journal}
  {\bibinfo  {journal} {Proceedings of the Physical Society. Section A}\
  }\textbf {\bibinfo {volume} {62}},\ \bibinfo {pages} {416} (\bibinfo {year}
  {1949})}\BibitemShut {NoStop}%
\bibitem [{\citenamefont {J\'erome}\ \emph {et~al.}(1967)\citenamefont
  {J\'erome}, \citenamefont {Rice},\ and\ \citenamefont
  {Kohn}}]{kohn_excitonic_ins}%
  \BibitemOpen
  \bibfield  {author} {\bibinfo {author} {\bibfnamefont {D.}~\bibnamefont
  {J\'erome}}, \bibinfo {author} {\bibfnamefont {T.~M.}\ \bibnamefont {Rice}},
  \ and\ \bibinfo {author} {\bibfnamefont {W.}~\bibnamefont {Kohn}},\ }\href
  {\doibase 10.1103/PhysRev.158.462} {\bibfield  {journal} {\bibinfo  {journal}
  {Phys. Rev.}\ }\textbf {\bibinfo {volume} {158}},\ \bibinfo {pages} {462}
  (\bibinfo {year} {1967})}\BibitemShut {NoStop}%
\bibitem [{\citenamefont {{Kozlov}}\ and\ \citenamefont
  {{Maksimov}}(1965)}]{kozlov_divalent_crystal}%
  \BibitemOpen
  \bibfield  {author} {\bibinfo {author} {\bibfnamefont {A.~N.}\ \bibnamefont
  {{Kozlov}}}\ and\ \bibinfo {author} {\bibfnamefont {L.~A.}\ \bibnamefont
  {{Maksimov}}},\ }\href@noop {} {\bibfield  {journal} {\bibinfo  {journal}
  {Soviet Journal of Experimental and Theoretical Physics}\ }\textbf {\bibinfo
  {volume} {21}},\ \bibinfo {pages} {790} (\bibinfo {year} {1965})}\BibitemShut
  {NoStop}%
\bibitem [{\citenamefont {Halperin}\ and\ \citenamefont
  {Rice}(1968)}]{halperin_rice_rmp}%
  \BibitemOpen
  \bibfield  {author} {\bibinfo {author} {\bibfnamefont {B.~I.}\ \bibnamefont
  {Halperin}}\ and\ \bibinfo {author} {\bibfnamefont {T.~M.}\ \bibnamefont
  {Rice}},\ }\href {\doibase 10.1103/RevModPhys.40.755} {\bibfield  {journal}
  {\bibinfo  {journal} {Rev. Mod. Phys.}\ }\textbf {\bibinfo {volume} {40}},\
  \bibinfo {pages} {755} (\bibinfo {year} {1968})}\BibitemShut {NoStop}%
\bibitem [{\citenamefont {{Keldysh}}\ and\ \citenamefont
  {{Kozlov}}(1968)}]{keldysh_collective_properties_excitons}%
  \BibitemOpen
  \bibfield  {author} {\bibinfo {author} {\bibfnamefont {L.~V.}\ \bibnamefont
  {{Keldysh}}}\ and\ \bibinfo {author} {\bibfnamefont {A.~N.}\ \bibnamefont
  {{Kozlov}}},\ }\href@noop {} {\bibfield  {journal} {\bibinfo  {journal}
  {Soviet Journal of Experimental and Theoretical Physics}\ }\textbf {\bibinfo
  {volume} {27}},\ \bibinfo {pages} {521} (\bibinfo {year} {1968})}\BibitemShut
  {NoStop}%
\bibitem [{\citenamefont {Spielman}\ \emph {et~al.}(2000)\citenamefont
  {Spielman}, \citenamefont {Eisenstein}, \citenamefont {Pfeiffer},\ and\
  \citenamefont {West}}]{spielman_double_quantum_hall_ferromagnet_prl2000}%
  \BibitemOpen
  \bibfield  {author} {\bibinfo {author} {\bibfnamefont {I.~B.}\ \bibnamefont
  {Spielman}}, \bibinfo {author} {\bibfnamefont {J.~P.}\ \bibnamefont
  {Eisenstein}}, \bibinfo {author} {\bibfnamefont {L.~N.}\ \bibnamefont
  {Pfeiffer}}, \ and\ \bibinfo {author} {\bibfnamefont {K.~W.}\ \bibnamefont
  {West}},\ }\href {\doibase 10.1103/PhysRevLett.84.5808} {\bibfield  {journal}
  {\bibinfo  {journal} {Phys. Rev. Lett.}\ }\textbf {\bibinfo {volume} {84}},\
  \bibinfo {pages} {5808} (\bibinfo {year} {2000})}\BibitemShut {NoStop}%
\bibitem [{\citenamefont {Kellogg}\ \emph {et~al.}(2004)\citenamefont
  {Kellogg}, \citenamefont {Eisenstein}, \citenamefont {Pfeiffer},\ and\
  \citenamefont {West}}]{kellogg_double_electron_gas_prl2004}%
  \BibitemOpen
  \bibfield  {author} {\bibinfo {author} {\bibfnamefont {M.}~\bibnamefont
  {Kellogg}}, \bibinfo {author} {\bibfnamefont {J.~P.}\ \bibnamefont
  {Eisenstein}}, \bibinfo {author} {\bibfnamefont {L.~N.}\ \bibnamefont
  {Pfeiffer}}, \ and\ \bibinfo {author} {\bibfnamefont {K.~W.}\ \bibnamefont
  {West}},\ }\href {\doibase 10.1103/PhysRevLett.93.036801} {\bibfield
  {journal} {\bibinfo  {journal} {Phys. Rev. Lett.}\ }\textbf {\bibinfo
  {volume} {93}},\ \bibinfo {pages} {036801} (\bibinfo {year}
  {2004})}\BibitemShut {NoStop}%
\bibitem [{\citenamefont {Eisenstein}\ and\ \citenamefont
  {MacDonald}(2004)}]{eisenstein_nature2004}%
  \BibitemOpen
  \bibfield  {author} {\bibinfo {author} {\bibfnamefont {J.~P.}\ \bibnamefont
  {Eisenstein}}\ and\ \bibinfo {author} {\bibfnamefont {A.~H.}\ \bibnamefont
  {MacDonald}},\ }\href {\doibase 10.1038/nature03081} {\bibfield  {journal}
  {\bibinfo  {journal} {Nature}\ }\textbf {\bibinfo {volume} {432}},\ \bibinfo
  {pages} {691} (\bibinfo {year} {2004})}\BibitemShut {NoStop}%
\bibitem [{\citenamefont {Eisenstein}(2014)}]{eisenstein_review_2014}%
  \BibitemOpen
  \bibfield  {author} {\bibinfo {author} {\bibfnamefont {J.}~\bibnamefont
  {Eisenstein}},\ }\href {\doibase 10.1146/annurev-conmatphys-031113-133832}
  {\bibfield  {journal} {\bibinfo  {journal} {Annual Review of Condensed Matter
  Physics}\ }\textbf {\bibinfo {volume} {5}},\ \bibinfo {pages} {159} (\bibinfo
  {year} {2014})},\ \Eprint
  {http://arxiv.org/abs/https://doi.org/10.1146/annurev-conmatphys-031113-133832}
  {https://doi.org/10.1146/annurev-conmatphys-031113-133832} \BibitemShut
  {NoStop}%
\bibitem [{\citenamefont {Ma}\ \emph {et~al.}(2021)\citenamefont {Ma},
  \citenamefont {Nguyen}, \citenamefont {Wang}, \citenamefont {Zeng},
  \citenamefont {Watanabe}, \citenamefont {Taniguchi}, \citenamefont
  {MacDonald}, \citenamefont {Mak},\ and\ \citenamefont
  {Shan}}]{excitonic_insulator_atomic_double_layers_Nat2021}%
  \BibitemOpen
  \bibfield  {author} {\bibinfo {author} {\bibfnamefont {L.}~\bibnamefont
  {Ma}}, \bibinfo {author} {\bibfnamefont {P.~X.}\ \bibnamefont {Nguyen}},
  \bibinfo {author} {\bibfnamefont {Z.}~\bibnamefont {Wang}}, \bibinfo {author}
  {\bibfnamefont {Y.}~\bibnamefont {Zeng}}, \bibinfo {author} {\bibfnamefont
  {K.}~\bibnamefont {Watanabe}}, \bibinfo {author} {\bibfnamefont
  {T.}~\bibnamefont {Taniguchi}}, \bibinfo {author} {\bibfnamefont {A.~H.}\
  \bibnamefont {MacDonald}}, \bibinfo {author} {\bibfnamefont {K.~F.}\
  \bibnamefont {Mak}}, \ and\ \bibinfo {author} {\bibfnamefont
  {J.}~\bibnamefont {Shan}},\ }\href {\doibase 10.1038/s41586-021-03947-9}
  {\bibfield  {journal} {\bibinfo  {journal} {Nature}\ }\textbf {\bibinfo
  {volume} {598}},\ \bibinfo {pages} {585} (\bibinfo {year}
  {2021})}\BibitemShut {NoStop}%
\bibitem [{\citenamefont {Nguyen}\ \emph {et~al.}(2025)\citenamefont {Nguyen},
  \citenamefont {Chaturvedi}, \citenamefont {Zou}, \citenamefont {Watanabe},
  \citenamefont {Taniguchi}, \citenamefont {MacDonald}, \citenamefont {Mak},\
  and\ \citenamefont {Shan}}]{quantum_oscillation_dipolar_excitons_NatMat2025}%
  \BibitemOpen
  \bibfield  {author} {\bibinfo {author} {\bibfnamefont {P.~X.}\ \bibnamefont
  {Nguyen}}, \bibinfo {author} {\bibfnamefont {R.}~\bibnamefont {Chaturvedi}},
  \bibinfo {author} {\bibfnamefont {B.}~\bibnamefont {Zou}}, \bibinfo {author}
  {\bibfnamefont {K.}~\bibnamefont {Watanabe}}, \bibinfo {author}
  {\bibfnamefont {T.}~\bibnamefont {Taniguchi}}, \bibinfo {author}
  {\bibfnamefont {A.~H.}\ \bibnamefont {MacDonald}}, \bibinfo {author}
  {\bibfnamefont {K.~F.}\ \bibnamefont {Mak}}, \ and\ \bibinfo {author}
  {\bibfnamefont {J.}~\bibnamefont {Shan}},\ }\href {\doibase
  10.1038/s41563-025-02334-3} {\bibfield  {journal} {\bibinfo  {journal}
  {Nature Materials}\ } (\bibinfo {year} {2025}),\
  10.1038/s41563-025-02334-3}\BibitemShut {NoStop}%
\bibitem [{\citenamefont {Lu}\ \emph {et~al.}(2017)\citenamefont {Lu},
  \citenamefont {Kono}, \citenamefont {Larkin}, \citenamefont {Rost},
  \citenamefont {Takayama}, \citenamefont {Boris}, \citenamefont {Keimer},\
  and\ \citenamefont {Takagi}}]{exp_ei_Ta2NiSe5}%
  \BibitemOpen
  \bibfield  {author} {\bibinfo {author} {\bibfnamefont {Y.~F.}\ \bibnamefont
  {Lu}}, \bibinfo {author} {\bibfnamefont {H.}~\bibnamefont {Kono}}, \bibinfo
  {author} {\bibfnamefont {T.~I.}\ \bibnamefont {Larkin}}, \bibinfo {author}
  {\bibfnamefont {A.~W.}\ \bibnamefont {Rost}}, \bibinfo {author}
  {\bibfnamefont {T.}~\bibnamefont {Takayama}}, \bibinfo {author}
  {\bibfnamefont {A.~V.}\ \bibnamefont {Boris}}, \bibinfo {author}
  {\bibfnamefont {B.}~\bibnamefont {Keimer}}, \ and\ \bibinfo {author}
  {\bibfnamefont {H.}~\bibnamefont {Takagi}},\ }\href
  {http://dx.doi.org/10.1038/ncomms14408} {\bibfield  {journal} {\bibinfo
  {journal} {Nat. Commun.}\ }\textbf {\bibinfo {volume} {8}},\ \bibinfo {pages}
  {14408} (\bibinfo {year} {2017})}\BibitemShut {NoStop}%
\bibitem [{\citenamefont {Kogar}\ \emph {et~al.}(2017)\citenamefont {Kogar},
  \citenamefont {Rak}, \citenamefont {Vig}, \citenamefont {Husain},
  \citenamefont {Flicker}, \citenamefont {Joe}, \citenamefont {Venema},
  \citenamefont {MacDougall}, \citenamefont {Chiang}, \citenamefont {Fradkin},
  \citenamefont {van Wezel},\ and\ \citenamefont {Abbamonte}}]{kogar_TiSe2}%
  \BibitemOpen
  \bibfield  {author} {\bibinfo {author} {\bibfnamefont {A.}~\bibnamefont
  {Kogar}}, \bibinfo {author} {\bibfnamefont {M.~S.}\ \bibnamefont {Rak}},
  \bibinfo {author} {\bibfnamefont {S.}~\bibnamefont {Vig}}, \bibinfo {author}
  {\bibfnamefont {A.~A.}\ \bibnamefont {Husain}}, \bibinfo {author}
  {\bibfnamefont {F.}~\bibnamefont {Flicker}}, \bibinfo {author} {\bibfnamefont
  {Y.~I.}\ \bibnamefont {Joe}}, \bibinfo {author} {\bibfnamefont
  {L.}~\bibnamefont {Venema}}, \bibinfo {author} {\bibfnamefont {G.~J.}\
  \bibnamefont {MacDougall}}, \bibinfo {author} {\bibfnamefont {T.~C.}\
  \bibnamefont {Chiang}}, \bibinfo {author} {\bibfnamefont {E.}~\bibnamefont
  {Fradkin}}, \bibinfo {author} {\bibfnamefont {J.}~\bibnamefont {van Wezel}},
  \ and\ \bibinfo {author} {\bibfnamefont {P.}~\bibnamefont {Abbamonte}},\
  }\href {\doibase 10.1126/science.aam6432} {\bibfield  {journal} {\bibinfo
  {journal} {Science}\ }\textbf {\bibinfo {volume} {358}},\ \bibinfo {pages}
  {1314} (\bibinfo {year} {2017})},\ \Eprint
  {http://arxiv.org/abs/http://science.sciencemag.org/content/358/6368/1314.full.pdf}
  {http://science.sciencemag.org/content/358/6368/1314.full.pdf} \BibitemShut
  {NoStop}%
\bibitem [{\citenamefont {Varsano}\ \emph {et~al.}(2020)\citenamefont
  {Varsano}, \citenamefont {Palummo}, \citenamefont {Molinari},\ and\
  \citenamefont {Rontani}}]{varsano_NatNano_2020}%
  \BibitemOpen
  \bibfield  {author} {\bibinfo {author} {\bibfnamefont {D.}~\bibnamefont
  {Varsano}}, \bibinfo {author} {\bibfnamefont {M.}~\bibnamefont {Palummo}},
  \bibinfo {author} {\bibfnamefont {E.}~\bibnamefont {Molinari}}, \ and\
  \bibinfo {author} {\bibfnamefont {M.}~\bibnamefont {Rontani}},\ }\href
  {\doibase 10.1038/s41565-020-0650-4} {\bibfield  {journal} {\bibinfo
  {journal} {Nature Nanotechnology}\ }\textbf {\bibinfo {volume} {15}},\
  \bibinfo {pages} {367} (\bibinfo {year} {2020})}\BibitemShut {NoStop}%
\bibitem [{\citenamefont {Ataei}\ \emph {et~al.}(2021)\citenamefont {Ataei},
  \citenamefont {Varsano}, \citenamefont {Molinari},\ and\ \citenamefont
  {Rontani}}]{varsano_pnas_2021}%
  \BibitemOpen
  \bibfield  {author} {\bibinfo {author} {\bibfnamefont {S.~S.}\ \bibnamefont
  {Ataei}}, \bibinfo {author} {\bibfnamefont {D.}~\bibnamefont {Varsano}},
  \bibinfo {author} {\bibfnamefont {E.}~\bibnamefont {Molinari}}, \ and\
  \bibinfo {author} {\bibfnamefont {M.}~\bibnamefont {Rontani}},\ }\href
  {\doibase 10.1073/pnas.2010110118} {\bibfield  {journal} {\bibinfo  {journal}
  {Proceedings of the National Academy of Sciences}\ }\textbf {\bibinfo
  {volume} {118}},\ \bibinfo {pages} {e2010110118} (\bibinfo {year} {2021})},\
  \Eprint
  {http://arxiv.org/abs/https://www.pnas.org/doi/pdf/10.1073/pnas.2010110118}
  {https://www.pnas.org/doi/pdf/10.1073/pnas.2010110118} \BibitemShut {NoStop}%
\bibitem [{\citenamefont {Jia}\ \emph {et~al.}(2022)\citenamefont {Jia},
  \citenamefont {Wang}, \citenamefont {Chiu}, \citenamefont {Song},
  \citenamefont {Yu}, \citenamefont {J{\"a}ck}, \citenamefont {Lei},
  \citenamefont {Klemenz}, \citenamefont {Cevallos}, \citenamefont {Onyszczak},
  \citenamefont {Fishchenko}, \citenamefont {Liu}, \citenamefont {Farahi},
  \citenamefont {Xie}, \citenamefont {Xu}, \citenamefont {Watanabe},
  \citenamefont {Taniguchi}, \citenamefont {Bernevig}, \citenamefont {Cava},
  \citenamefont {Schoop}, \citenamefont {Yazdani},\ and\ \citenamefont
  {Wu}}]{ei_monolayer_wte2_NatPhys_2022}%
  \BibitemOpen
  \bibfield  {author} {\bibinfo {author} {\bibfnamefont {Y.}~\bibnamefont
  {Jia}}, \bibinfo {author} {\bibfnamefont {P.}~\bibnamefont {Wang}}, \bibinfo
  {author} {\bibfnamefont {C.-L.}\ \bibnamefont {Chiu}}, \bibinfo {author}
  {\bibfnamefont {Z.}~\bibnamefont {Song}}, \bibinfo {author} {\bibfnamefont
  {G.}~\bibnamefont {Yu}}, \bibinfo {author} {\bibfnamefont {B.}~\bibnamefont
  {J{\"a}ck}}, \bibinfo {author} {\bibfnamefont {S.}~\bibnamefont {Lei}},
  \bibinfo {author} {\bibfnamefont {S.}~\bibnamefont {Klemenz}}, \bibinfo
  {author} {\bibfnamefont {F.~A.}\ \bibnamefont {Cevallos}}, \bibinfo {author}
  {\bibfnamefont {M.}~\bibnamefont {Onyszczak}}, \bibinfo {author}
  {\bibfnamefont {N.}~\bibnamefont {Fishchenko}}, \bibinfo {author}
  {\bibfnamefont {X.}~\bibnamefont {Liu}}, \bibinfo {author} {\bibfnamefont
  {G.}~\bibnamefont {Farahi}}, \bibinfo {author} {\bibfnamefont
  {F.}~\bibnamefont {Xie}}, \bibinfo {author} {\bibfnamefont {Y.}~\bibnamefont
  {Xu}}, \bibinfo {author} {\bibfnamefont {K.}~\bibnamefont {Watanabe}},
  \bibinfo {author} {\bibfnamefont {T.}~\bibnamefont {Taniguchi}}, \bibinfo
  {author} {\bibfnamefont {B.~A.}\ \bibnamefont {Bernevig}}, \bibinfo {author}
  {\bibfnamefont {R.~J.}\ \bibnamefont {Cava}}, \bibinfo {author}
  {\bibfnamefont {L.~M.}\ \bibnamefont {Schoop}}, \bibinfo {author}
  {\bibfnamefont {A.}~\bibnamefont {Yazdani}}, \ and\ \bibinfo {author}
  {\bibfnamefont {S.}~\bibnamefont {Wu}},\ }\href {\doibase
  10.1038/s41567-021-01422-w} {\bibfield  {journal} {\bibinfo  {journal}
  {Nature Physics}\ }\textbf {\bibinfo {volume} {18}},\ \bibinfo {pages} {87}
  (\bibinfo {year} {2022})}\BibitemShut {NoStop}%
\bibitem [{\citenamefont {Sun}\ \emph {et~al.}(2022)\citenamefont {Sun},
  \citenamefont {Zhao}, \citenamefont {Palomaki}, \citenamefont {Fei},
  \citenamefont {Runburg}, \citenamefont {Malinowski}, \citenamefont {Huang},
  \citenamefont {Cenker}, \citenamefont {Cui}, \citenamefont {Chu},
  \citenamefont {Xu}, \citenamefont {Ataei}, \citenamefont {Varsano},
  \citenamefont {Palummo}, \citenamefont {Molinari}, \citenamefont {Rontani},\
  and\ \citenamefont {Cobden}}]{EI_wte2_NatPhys2022}%
  \BibitemOpen
  \bibfield  {author} {\bibinfo {author} {\bibfnamefont {B.}~\bibnamefont
  {Sun}}, \bibinfo {author} {\bibfnamefont {W.}~\bibnamefont {Zhao}}, \bibinfo
  {author} {\bibfnamefont {T.}~\bibnamefont {Palomaki}}, \bibinfo {author}
  {\bibfnamefont {Z.}~\bibnamefont {Fei}}, \bibinfo {author} {\bibfnamefont
  {E.}~\bibnamefont {Runburg}}, \bibinfo {author} {\bibfnamefont
  {P.}~\bibnamefont {Malinowski}}, \bibinfo {author} {\bibfnamefont
  {X.}~\bibnamefont {Huang}}, \bibinfo {author} {\bibfnamefont
  {J.}~\bibnamefont {Cenker}}, \bibinfo {author} {\bibfnamefont {Y.-T.}\
  \bibnamefont {Cui}}, \bibinfo {author} {\bibfnamefont {J.-H.}\ \bibnamefont
  {Chu}}, \bibinfo {author} {\bibfnamefont {X.}~\bibnamefont {Xu}}, \bibinfo
  {author} {\bibfnamefont {S.~S.}\ \bibnamefont {Ataei}}, \bibinfo {author}
  {\bibfnamefont {D.}~\bibnamefont {Varsano}}, \bibinfo {author} {\bibfnamefont
  {M.}~\bibnamefont {Palummo}}, \bibinfo {author} {\bibfnamefont
  {E.}~\bibnamefont {Molinari}}, \bibinfo {author} {\bibfnamefont
  {M.}~\bibnamefont {Rontani}}, \ and\ \bibinfo {author} {\bibfnamefont
  {D.~H.}\ \bibnamefont {Cobden}},\ }\href {\doibase
  10.1038/s41567-021-01427-5} {\bibfield  {journal} {\bibinfo  {journal}
  {Nature Physics}\ }\textbf {\bibinfo {volume} {18}},\ \bibinfo {pages} {94}
  (\bibinfo {year} {2022})}\BibitemShut {NoStop}%
\bibitem [{\citenamefont {Mei}\ \emph {et~al.}(2025)\citenamefont {Mei},
  \citenamefont {Wang}, \citenamefont {Fei}, \citenamefont {Wang},
  \citenamefont {Gan}, \citenamefont {Liu},\ and\ \citenamefont
  {Wang}}]{ei_black_phosphorous_2025}%
  \BibitemOpen
  \bibfield  {author} {\bibinfo {author} {\bibfnamefont {J.}~\bibnamefont
  {Mei}}, \bibinfo {author} {\bibfnamefont {Y.}~\bibnamefont {Wang}}, \bibinfo
  {author} {\bibfnamefont {R.}~\bibnamefont {Fei}}, \bibinfo {author}
  {\bibfnamefont {J.}~\bibnamefont {Wang}}, \bibinfo {author} {\bibfnamefont
  {X.}~\bibnamefont {Gan}}, \bibinfo {author} {\bibfnamefont {B.}~\bibnamefont
  {Liu}}, \ and\ \bibinfo {author} {\bibfnamefont {X.}~\bibnamefont {Wang}},\
  }\href {\doibase 10.1038/s41467-025-58886-0} {\bibfield  {journal} {\bibinfo
  {journal} {Nature Communications}\ }\textbf {\bibinfo {volume} {16}},\
  \bibinfo {pages} {3744} (\bibinfo {year} {2025})}\BibitemShut {NoStop}%
\bibitem [{\citenamefont {Hossain}\ \emph {et~al.}(2025)\citenamefont
  {Hossain}, \citenamefont {Cheng}, \citenamefont {Jiang}, \citenamefont
  {Cochran}, \citenamefont {Zhang}, \citenamefont {Wu}, \citenamefont {Liu},
  \citenamefont {Zheng}, \citenamefont {Cheng}, \citenamefont {Kim},
  \citenamefont {Zhang}, \citenamefont {Litskevich}, \citenamefont {Zhang},
  \citenamefont {Liu}, \citenamefont {Yin}, \citenamefont {Yang}, \citenamefont
  {Denlinger}, \citenamefont {Tallarida}, \citenamefont {Dai}, \citenamefont
  {Vescovo}, \citenamefont {Rajapitamahuni}, \citenamefont {Yao}, \citenamefont
  {Keselman}, \citenamefont {Peng}, \citenamefont {Yao}, \citenamefont {Wang},
  \citenamefont {Balicas}, \citenamefont {Neupert},\ and\ \citenamefont
  {Hasan}}]{hossainTPT_nat_phys_2025}%
  \BibitemOpen
  \bibfield  {author} {\bibinfo {author} {\bibfnamefont {M.~S.}\ \bibnamefont
  {Hossain}}, \bibinfo {author} {\bibfnamefont {Z.-J.}\ \bibnamefont {Cheng}},
  \bibinfo {author} {\bibfnamefont {Y.-X.}\ \bibnamefont {Jiang}}, \bibinfo
  {author} {\bibfnamefont {T.~A.}\ \bibnamefont {Cochran}}, \bibinfo {author}
  {\bibfnamefont {S.-B.}\ \bibnamefont {Zhang}}, \bibinfo {author}
  {\bibfnamefont {H.}~\bibnamefont {Wu}}, \bibinfo {author} {\bibfnamefont
  {X.}~\bibnamefont {Liu}}, \bibinfo {author} {\bibfnamefont {X.}~\bibnamefont
  {Zheng}}, \bibinfo {author} {\bibfnamefont {G.}~\bibnamefont {Cheng}},
  \bibinfo {author} {\bibfnamefont {B.}~\bibnamefont {Kim}}, \bibinfo {author}
  {\bibfnamefont {Q.}~\bibnamefont {Zhang}}, \bibinfo {author} {\bibfnamefont
  {M.}~\bibnamefont {Litskevich}}, \bibinfo {author} {\bibfnamefont
  {J.}~\bibnamefont {Zhang}}, \bibinfo {author} {\bibfnamefont
  {J.}~\bibnamefont {Liu}}, \bibinfo {author} {\bibfnamefont {J.-X.}\
  \bibnamefont {Yin}}, \bibinfo {author} {\bibfnamefont {X.~P.}\ \bibnamefont
  {Yang}}, \bibinfo {author} {\bibfnamefont {J.~D.}\ \bibnamefont {Denlinger}},
  \bibinfo {author} {\bibfnamefont {M.}~\bibnamefont {Tallarida}}, \bibinfo
  {author} {\bibfnamefont {J.}~\bibnamefont {Dai}}, \bibinfo {author}
  {\bibfnamefont {E.}~\bibnamefont {Vescovo}}, \bibinfo {author} {\bibfnamefont
  {A.}~\bibnamefont {Rajapitamahuni}}, \bibinfo {author} {\bibfnamefont
  {N.}~\bibnamefont {Yao}}, \bibinfo {author} {\bibfnamefont {A.}~\bibnamefont
  {Keselman}}, \bibinfo {author} {\bibfnamefont {Y.}~\bibnamefont {Peng}},
  \bibinfo {author} {\bibfnamefont {Y.}~\bibnamefont {Yao}}, \bibinfo {author}
  {\bibfnamefont {Z.}~\bibnamefont {Wang}}, \bibinfo {author} {\bibfnamefont
  {L.}~\bibnamefont {Balicas}}, \bibinfo {author} {\bibfnamefont
  {T.}~\bibnamefont {Neupert}}, \ and\ \bibinfo {author} {\bibfnamefont
  {M.~Z.}\ \bibnamefont {Hasan}},\ }\href {\doibase 10.1038/s41567-025-02917-6}
  {\bibfield  {journal} {\bibinfo  {journal} {Nature Physics}\ }\textbf
  {\bibinfo {volume} {21}},\ \bibinfo {pages} {1250} (\bibinfo {year}
  {2025})}\BibitemShut {NoStop}%
\bibitem [{\citenamefont {Mazza}\ and\ \citenamefont
  {Polini}(2023)}]{giacomo_hiddenTRSB2023}%
  \BibitemOpen
  \bibfield  {author} {\bibinfo {author} {\bibfnamefont {G.}~\bibnamefont
  {Mazza}}\ and\ \bibinfo {author} {\bibfnamefont {M.}~\bibnamefont {Polini}},\
  }\href {\doibase 10.1103/PhysRevB.108.L241107} {\bibfield  {journal}
  {\bibinfo  {journal} {Phys. Rev. B}\ }\textbf {\bibinfo {volume} {108}},\
  \bibinfo {pages} {L241107} (\bibinfo {year} {2023})}\BibitemShut {NoStop}%
\bibitem [{\citenamefont {Wang}\ \emph {et~al.}(2023)\citenamefont {Wang},
  \citenamefont {Papaj},\ and\ \citenamefont
  {Moore}}]{breakdown_topological_TRSB_papaj_2023}%
  \BibitemOpen
  \bibfield  {author} {\bibinfo {author} {\bibfnamefont {Y.-Q.}\ \bibnamefont
  {Wang}}, \bibinfo {author} {\bibfnamefont {M.}~\bibnamefont {Papaj}}, \ and\
  \bibinfo {author} {\bibfnamefont {J.~E.}\ \bibnamefont {Moore}},\ }\href
  {\doibase 10.1103/PhysRevB.108.205420} {\bibfield  {journal} {\bibinfo
  {journal} {Phys. Rev. B}\ }\textbf {\bibinfo {volume} {108}},\ \bibinfo
  {pages} {205420} (\bibinfo {year} {2023})}\BibitemShut {NoStop}%
\bibitem [{\citenamefont {Amaricci}\ \emph {et~al.}(2023)\citenamefont
  {Amaricci}, \citenamefont {Mazza}, \citenamefont {Capone},\ and\
  \citenamefont {Fabrizio}}]{amaricci_QSHI2023}%
  \BibitemOpen
  \bibfield  {author} {\bibinfo {author} {\bibfnamefont {A.}~\bibnamefont
  {Amaricci}}, \bibinfo {author} {\bibfnamefont {G.}~\bibnamefont {Mazza}},
  \bibinfo {author} {\bibfnamefont {M.}~\bibnamefont {Capone}}, \ and\ \bibinfo
  {author} {\bibfnamefont {M.}~\bibnamefont {Fabrizio}},\ }\href {\doibase
  10.1103/PhysRevB.107.115117} {\bibfield  {journal} {\bibinfo  {journal}
  {Phys. Rev. B}\ }\textbf {\bibinfo {volume} {107}},\ \bibinfo {pages}
  {115117} (\bibinfo {year} {2023})}\BibitemShut {NoStop}%
\bibitem [{\citenamefont
  {Papaj}(2024)}]{spectroscopic_signatures_papaj_prb2024}%
  \BibitemOpen
  \bibfield  {author} {\bibinfo {author} {\bibfnamefont {M.}~\bibnamefont
  {Papaj}},\ }\href {\doibase 10.1103/PhysRevB.110.165422} {\bibfield
  {journal} {\bibinfo  {journal} {Phys. Rev. B}\ }\textbf {\bibinfo {volume}
  {110}},\ \bibinfo {pages} {165422} (\bibinfo {year} {2024})}\BibitemShut
  {NoStop}%
\bibitem [{\citenamefont {Phan}(2025)}]{Van-Nham_2025}%
  \BibitemOpen
  \bibfield  {author} {\bibinfo {author} {\bibfnamefont {V.-N.}\ \bibnamefont
  {Phan}},\ }\href {\doibase 10.1103/7lp7-p8tk} {\bibfield  {journal} {\bibinfo
   {journal} {Phys. Rev. B}\ }\textbf {\bibinfo {volume} {112}},\ \bibinfo
  {pages} {L201110} (\bibinfo {year} {2025})}\BibitemShut {NoStop}%
\bibitem [{\citenamefont {Ohta}\ and\ \citenamefont
  {Nasu}(2025)}]{nasu_magnetic_field_effect}%
  \BibitemOpen
  \bibfield  {author} {\bibinfo {author} {\bibfnamefont {N.}~\bibnamefont
  {Ohta}}\ and\ \bibinfo {author} {\bibfnamefont {J.}~\bibnamefont {Nasu}},\
  }\href {\doibase 10.7566/JPSJ.94.054702} {\bibfield  {journal} {\bibinfo
  {journal} {Journal of the Physical Society of Japan}\ }\textbf {\bibinfo
  {volume} {94}},\ \bibinfo {pages} {054702} (\bibinfo {year} {2025})},\
  \Eprint {http://arxiv.org/abs/https://doi.org/10.7566/JPSJ.94.054702}
  {https://doi.org/10.7566/JPSJ.94.054702} \BibitemShut {NoStop}%
\bibitem [{\citenamefont {Salvo}\ \emph {et~al.}(1986)\citenamefont {Salvo},
  \citenamefont {Chen}, \citenamefont {Fleming}, \citenamefont {Waszczak},
  \citenamefont {Dunn}, \citenamefont {Sunshine},\ and\ \citenamefont
  {Ibers}}]{disalvo_Ta2NiSe5}%
  \BibitemOpen
  \bibfield  {author} {\bibinfo {author} {\bibfnamefont {F.~D.}\ \bibnamefont
  {Salvo}}, \bibinfo {author} {\bibfnamefont {C.}~\bibnamefont {Chen}},
  \bibinfo {author} {\bibfnamefont {R.}~\bibnamefont {Fleming}}, \bibinfo
  {author} {\bibfnamefont {J.}~\bibnamefont {Waszczak}}, \bibinfo {author}
  {\bibfnamefont {R.}~\bibnamefont {Dunn}}, \bibinfo {author} {\bibfnamefont
  {S.}~\bibnamefont {Sunshine}}, \ and\ \bibinfo {author} {\bibfnamefont
  {J.~A.}\ \bibnamefont {Ibers}},\ }\href {\doibase
  https://doi.org/10.1016/0022-5088(86)90216-X} {\bibfield  {journal} {\bibinfo
   {journal} {Journal of the Less Common Metals}\ }\textbf {\bibinfo {volume}
  {116}},\ \bibinfo {pages} {51 } (\bibinfo {year} {1986})}\BibitemShut
  {NoStop}%
\bibitem [{\citenamefont {Seki}\ \emph {et~al.}(2014)\citenamefont {Seki},
  \citenamefont {Wakisaka}, \citenamefont {Kaneko}, \citenamefont {Toriyama},
  \citenamefont {Konishi}, \citenamefont {Sudayama}, \citenamefont {Saini},
  \citenamefont {Arita}, \citenamefont {Namatame}, \citenamefont {Taniguchi},
  \citenamefont {Katayama}, \citenamefont {Nohara}, \citenamefont {Takagi},
  \citenamefont {Mizokawa},\ and\ \citenamefont {Ohta}}]{seki_TNS}%
  \BibitemOpen
  \bibfield  {author} {\bibinfo {author} {\bibfnamefont {K.}~\bibnamefont
  {Seki}}, \bibinfo {author} {\bibfnamefont {Y.}~\bibnamefont {Wakisaka}},
  \bibinfo {author} {\bibfnamefont {T.}~\bibnamefont {Kaneko}}, \bibinfo
  {author} {\bibfnamefont {T.}~\bibnamefont {Toriyama}}, \bibinfo {author}
  {\bibfnamefont {T.}~\bibnamefont {Konishi}}, \bibinfo {author} {\bibfnamefont
  {T.}~\bibnamefont {Sudayama}}, \bibinfo {author} {\bibfnamefont {N.~L.}\
  \bibnamefont {Saini}}, \bibinfo {author} {\bibfnamefont {M.}~\bibnamefont
  {Arita}}, \bibinfo {author} {\bibfnamefont {H.}~\bibnamefont {Namatame}},
  \bibinfo {author} {\bibfnamefont {M.}~\bibnamefont {Taniguchi}}, \bibinfo
  {author} {\bibfnamefont {N.}~\bibnamefont {Katayama}}, \bibinfo {author}
  {\bibfnamefont {M.}~\bibnamefont {Nohara}}, \bibinfo {author} {\bibfnamefont
  {H.}~\bibnamefont {Takagi}}, \bibinfo {author} {\bibfnamefont
  {T.}~\bibnamefont {Mizokawa}}, \ and\ \bibinfo {author} {\bibfnamefont
  {Y.}~\bibnamefont {Ohta}},\ }\href {\doibase 10.1103/PhysRevB.90.155116}
  {\bibfield  {journal} {\bibinfo  {journal} {Phys. Rev. B}\ }\textbf {\bibinfo
  {volume} {90}},\ \bibinfo {pages} {155116} (\bibinfo {year}
  {2014})}\BibitemShut {NoStop}%
\bibitem [{\citenamefont {Mazza}\ \emph {et~al.}(2020)\citenamefont {Mazza},
  \citenamefont {R\"osner}, \citenamefont {Windg\"atter}, \citenamefont
  {Latini}, \citenamefont {H\"ubener}, \citenamefont {Millis}, \citenamefont
  {Rubio},\ and\ \citenamefont {Georges}}]{giacomo_TNS}%
  \BibitemOpen
  \bibfield  {author} {\bibinfo {author} {\bibfnamefont {G.}~\bibnamefont
  {Mazza}}, \bibinfo {author} {\bibfnamefont {M.}~\bibnamefont {R\"osner}},
  \bibinfo {author} {\bibfnamefont {L.}~\bibnamefont {Windg\"atter}}, \bibinfo
  {author} {\bibfnamefont {S.}~\bibnamefont {Latini}}, \bibinfo {author}
  {\bibfnamefont {H.}~\bibnamefont {H\"ubener}}, \bibinfo {author}
  {\bibfnamefont {A.~J.}\ \bibnamefont {Millis}}, \bibinfo {author}
  {\bibfnamefont {A.}~\bibnamefont {Rubio}}, \ and\ \bibinfo {author}
  {\bibfnamefont {A.}~\bibnamefont {Georges}},\ }\href {\doibase
  10.1103/PhysRevLett.124.197601} {\bibfield  {journal} {\bibinfo  {journal}
  {Phys. Rev. Lett.}\ }\textbf {\bibinfo {volume} {124}},\ \bibinfo {pages}
  {197601} (\bibinfo {year} {2020})}\BibitemShut {NoStop}%
\bibitem [{\citenamefont {Kaneko}\ \emph {et~al.}(2013)\citenamefont {Kaneko},
  \citenamefont {Toriyama}, \citenamefont {Konishi},\ and\ \citenamefont
  {Ohta}}]{kaneko_ortho_to_mono}%
  \BibitemOpen
  \bibfield  {author} {\bibinfo {author} {\bibfnamefont {T.}~\bibnamefont
  {Kaneko}}, \bibinfo {author} {\bibfnamefont {T.}~\bibnamefont {Toriyama}},
  \bibinfo {author} {\bibfnamefont {T.}~\bibnamefont {Konishi}}, \ and\
  \bibinfo {author} {\bibfnamefont {Y.}~\bibnamefont {Ohta}},\ }\href {\doibase
  10.1103/PhysRevB.87.035121} {\bibfield  {journal} {\bibinfo  {journal} {Phys.
  Rev. B}\ }\textbf {\bibinfo {volume} {87}},\ \bibinfo {pages} {035121}
  (\bibinfo {year} {2013})}\BibitemShut {NoStop}%
\bibitem [{\citenamefont {Watson}\ \emph {et~al.}(2020)\citenamefont {Watson},
  \citenamefont {Markovi\ifmmode~\acute{c}\else \'{c}\fi{}}, \citenamefont
  {Morales}, \citenamefont {Le~F\`evre}, \citenamefont {Merz}, \citenamefont
  {Haghighirad},\ and\ \citenamefont {King}}]{watson_tns2020}%
  \BibitemOpen
  \bibfield  {author} {\bibinfo {author} {\bibfnamefont {M.~D.}\ \bibnamefont
  {Watson}}, \bibinfo {author} {\bibfnamefont {I.}~\bibnamefont
  {Markovi\ifmmode~\acute{c}\else \'{c}\fi{}}}, \bibinfo {author}
  {\bibfnamefont {E.~A.}\ \bibnamefont {Morales}}, \bibinfo {author}
  {\bibfnamefont {P.}~\bibnamefont {Le~F\`evre}}, \bibinfo {author}
  {\bibfnamefont {M.}~\bibnamefont {Merz}}, \bibinfo {author} {\bibfnamefont
  {A.~A.}\ \bibnamefont {Haghighirad}}, \ and\ \bibinfo {author} {\bibfnamefont
  {P.~D.~C.}\ \bibnamefont {King}},\ }\href {\doibase
  10.1103/PhysRevResearch.2.013236} {\bibfield  {journal} {\bibinfo  {journal}
  {Phys. Rev. Research}\ }\textbf {\bibinfo {volume} {2}},\ \bibinfo {pages}
  {013236} (\bibinfo {year} {2020})}\BibitemShut {NoStop}%
\bibitem [{\citenamefont {Fujimori}(2020)}]{journal_club_chicken_egg}%
  \BibitemOpen
  \bibfield  {author} {\bibinfo {author} {\bibfnamefont {A.}~\bibnamefont
  {Fujimori}},\ }\href {\doibase 10.36471/JCCM\_August\_2020\_01} {\bibfield
  {journal} {\bibinfo  {journal} {Journal Club for Condensed Matter Physics}\ }
  (\bibinfo {year} {2020}),\ 10.36471/JCCM\_August\_2020\_01}\BibitemShut
  {NoStop}%
\bibitem [{\citenamefont {Werdehausen}\ \emph {et~al.}(2018)\citenamefont
  {Werdehausen}, \citenamefont {Takayama}, \citenamefont {H{\"o}ppner},
  \citenamefont {Albrecht}, \citenamefont {Rost}, \citenamefont {Lu},
  \citenamefont {Manske}, \citenamefont {Takagi},\ and\ \citenamefont
  {Kaiser}}]{kaiser_TNS}%
  \BibitemOpen
  \bibfield  {author} {\bibinfo {author} {\bibfnamefont {D.}~\bibnamefont
  {Werdehausen}}, \bibinfo {author} {\bibfnamefont {T.}~\bibnamefont
  {Takayama}}, \bibinfo {author} {\bibfnamefont {M.}~\bibnamefont
  {H{\"o}ppner}}, \bibinfo {author} {\bibfnamefont {G.}~\bibnamefont
  {Albrecht}}, \bibinfo {author} {\bibfnamefont {A.~W.}\ \bibnamefont {Rost}},
  \bibinfo {author} {\bibfnamefont {Y.}~\bibnamefont {Lu}}, \bibinfo {author}
  {\bibfnamefont {D.}~\bibnamefont {Manske}}, \bibinfo {author} {\bibfnamefont
  {H.}~\bibnamefont {Takagi}}, \ and\ \bibinfo {author} {\bibfnamefont
  {S.}~\bibnamefont {Kaiser}},\ }\href {\doibase 10.1126/sciadv.aap8652}
  {\bibfield  {journal} {\bibinfo  {journal} {Science Advances}\ }\textbf
  {\bibinfo {volume} {4}} (\bibinfo {year} {2018}),\ 10.1126/sciadv.aap8652},\
  \Eprint
  {http://arxiv.org/abs/http://advances.sciencemag.org/content/4/3/eaap8652.full.pdf}
  {http://advances.sciencemag.org/content/4/3/eaap8652.full.pdf} \BibitemShut
  {NoStop}%
\bibitem [{\citenamefont {Gole\ifmmode~\check{z}\else \v{z}\fi{}}\ \emph
  {et~al.}(2020)\citenamefont {Gole\ifmmode~\check{z}\else \v{z}\fi{}},
  \citenamefont {Sun}, \citenamefont {Murakami}, \citenamefont {Georges},\ and\
  \citenamefont {Millis}}]{golez_nonlinear_spectroscopy_prl2020}%
  \BibitemOpen
  \bibfield  {author} {\bibinfo {author} {\bibfnamefont {D.}~\bibnamefont
  {Gole\ifmmode~\check{z}\else \v{z}\fi{}}}, \bibinfo {author} {\bibfnamefont
  {Z.}~\bibnamefont {Sun}}, \bibinfo {author} {\bibfnamefont {Y.}~\bibnamefont
  {Murakami}}, \bibinfo {author} {\bibfnamefont {A.}~\bibnamefont {Georges}}, \
  and\ \bibinfo {author} {\bibfnamefont {A.~J.}\ \bibnamefont {Millis}},\
  }\href {\doibase 10.1103/PhysRevLett.125.257601} {\bibfield  {journal}
  {\bibinfo  {journal} {Phys. Rev. Lett.}\ }\textbf {\bibinfo {volume} {125}},\
  \bibinfo {pages} {257601} (\bibinfo {year} {2020})}\BibitemShut {NoStop}%
\bibitem [{\citenamefont {Kim}\ \emph {et~al.}(2021)\citenamefont {Kim},
  \citenamefont {Kim}, \citenamefont {Kim}, \citenamefont {Kwon}, \citenamefont
  {Kim},\ and\ \citenamefont {Kim}}]{Kim2021_raman_TNS}%
  \BibitemOpen
  \bibfield  {author} {\bibinfo {author} {\bibfnamefont {K.}~\bibnamefont
  {Kim}}, \bibinfo {author} {\bibfnamefont {H.}~\bibnamefont {Kim}}, \bibinfo
  {author} {\bibfnamefont {J.}~\bibnamefont {Kim}}, \bibinfo {author}
  {\bibfnamefont {C.}~\bibnamefont {Kwon}}, \bibinfo {author} {\bibfnamefont
  {J.~S.}\ \bibnamefont {Kim}}, \ and\ \bibinfo {author} {\bibfnamefont
  {B.~J.}\ \bibnamefont {Kim}},\ }\href {\doibase 10.1038/s41467-021-22133-z}
  {\bibfield  {journal} {\bibinfo  {journal} {Nature Communications}\ }\textbf
  {\bibinfo {volume} {12}},\ \bibinfo {pages} {1969} (\bibinfo {year}
  {2021})}\BibitemShut {NoStop}%
\bibitem [{\citenamefont {Ye}\ \emph {et~al.}(2021)\citenamefont {Ye},
  \citenamefont {Volkov}, \citenamefont {Lohani}, \citenamefont {Feldman},
  \citenamefont {Kim}, \citenamefont {Kanigel},\ and\ \citenamefont
  {Blumberg}}]{Ye2021_raman_TNS}%
  \BibitemOpen
  \bibfield  {author} {\bibinfo {author} {\bibfnamefont {M.}~\bibnamefont
  {Ye}}, \bibinfo {author} {\bibfnamefont {P.~A.}\ \bibnamefont {Volkov}},
  \bibinfo {author} {\bibfnamefont {H.}~\bibnamefont {Lohani}}, \bibinfo
  {author} {\bibfnamefont {I.}~\bibnamefont {Feldman}}, \bibinfo {author}
  {\bibfnamefont {M.}~\bibnamefont {Kim}}, \bibinfo {author} {\bibfnamefont
  {A.}~\bibnamefont {Kanigel}}, \ and\ \bibinfo {author} {\bibfnamefont
  {G.}~\bibnamefont {Blumberg}},\ }\href {\doibase 10.1103/PhysRevB.104.045102}
  {\bibfield  {journal} {\bibinfo  {journal} {Phys. Rev. B}\ }\textbf {\bibinfo
  {volume} {104}},\ \bibinfo {pages} {045102} (\bibinfo {year}
  {2021})}\BibitemShut {NoStop}%
\bibitem [{\citenamefont {Subedi}(2020)}]{subedi_TNS}%
  \BibitemOpen
  \bibfield  {author} {\bibinfo {author} {\bibfnamefont {A.}~\bibnamefont
  {Subedi}},\ }\href {\doibase 10.1103/PhysRevMaterials.4.083601} {\bibfield
  {journal} {\bibinfo  {journal} {Phys. Rev. Mater.}\ }\textbf {\bibinfo
  {volume} {4}},\ \bibinfo {pages} {083601} (\bibinfo {year}
  {2020})}\BibitemShut {NoStop}%
\bibitem [{\citenamefont {Windg{\"a}tter}\ \emph {et~al.}(2021)\citenamefont
  {Windg{\"a}tter}, \citenamefont {R{\"o}sner}, \citenamefont {Mazza},
  \citenamefont {H{\"u}bener}, \citenamefont {Georges}, \citenamefont {Millis},
  \citenamefont {Latini},\ and\ \citenamefont {Rubio}}]{lukas_TNS}%
  \BibitemOpen
  \bibfield  {author} {\bibinfo {author} {\bibfnamefont {L.}~\bibnamefont
  {Windg{\"a}tter}}, \bibinfo {author} {\bibfnamefont {M.}~\bibnamefont
  {R{\"o}sner}}, \bibinfo {author} {\bibfnamefont {G.}~\bibnamefont {Mazza}},
  \bibinfo {author} {\bibfnamefont {H.}~\bibnamefont {H{\"u}bener}}, \bibinfo
  {author} {\bibfnamefont {A.}~\bibnamefont {Georges}}, \bibinfo {author}
  {\bibfnamefont {A.~J.}\ \bibnamefont {Millis}}, \bibinfo {author}
  {\bibfnamefont {S.}~\bibnamefont {Latini}}, \ and\ \bibinfo {author}
  {\bibfnamefont {A.}~\bibnamefont {Rubio}},\ }\href {\doibase
  10.1038/s41524-021-00675-6} {\bibfield  {journal} {\bibinfo  {journal} {npj
  Computational Materials}\ }\textbf {\bibinfo {volume} {7}},\ \bibinfo {pages}
  {210} (\bibinfo {year} {2021})}\BibitemShut {NoStop}%
\bibitem [{\citenamefont {Volkov}\ \emph {et~al.}(2021)\citenamefont {Volkov},
  \citenamefont {Ye}, \citenamefont {Lohani}, \citenamefont {Feldman},
  \citenamefont {Kanigel},\ and\ \citenamefont {Blumberg}}]{volkov_npjQM_2021}%
  \BibitemOpen
  \bibfield  {author} {\bibinfo {author} {\bibfnamefont {P.~A.}\ \bibnamefont
  {Volkov}}, \bibinfo {author} {\bibfnamefont {M.}~\bibnamefont {Ye}}, \bibinfo
  {author} {\bibfnamefont {H.}~\bibnamefont {Lohani}}, \bibinfo {author}
  {\bibfnamefont {I.}~\bibnamefont {Feldman}}, \bibinfo {author} {\bibfnamefont
  {A.}~\bibnamefont {Kanigel}}, \ and\ \bibinfo {author} {\bibfnamefont
  {G.}~\bibnamefont {Blumberg}},\ }\href {\doibase 10.1038/s41535-021-00351-4}
  {\bibfield  {journal} {\bibinfo  {journal} {npj Quantum Materials}\ }\textbf
  {\bibinfo {volume} {6}},\ \bibinfo {pages} {52} (\bibinfo {year}
  {2021})}\BibitemShut {NoStop}%
\bibitem [{\citenamefont {Gole\ifmmode~\check{z}\else \v{z}\fi{}}\ \emph
  {et~al.}(2022)\citenamefont {Gole\ifmmode~\check{z}\else \v{z}\fi{}},
  \citenamefont {Dufresne}, \citenamefont {Kim}, \citenamefont {Boschini},
  \citenamefont {Chu}, \citenamefont {Murakami}, \citenamefont {Levy},
  \citenamefont {Mills}, \citenamefont {Zhdanovich}, \citenamefont {Isobe},
  \citenamefont {Takagi}, \citenamefont {Kaiser}, \citenamefont {Werner},
  \citenamefont {Jones}, \citenamefont {Georges}, \citenamefont {Damascelli},\
  and\ \citenamefont {Millis}}]{golez_unveiling_2022}%
  \BibitemOpen
  \bibfield  {author} {\bibinfo {author} {\bibfnamefont {D.}~\bibnamefont
  {Gole\ifmmode~\check{z}\else \v{z}\fi{}}}, \bibinfo {author} {\bibfnamefont
  {S.~K.~Y.}\ \bibnamefont {Dufresne}}, \bibinfo {author} {\bibfnamefont
  {M.-J.}\ \bibnamefont {Kim}}, \bibinfo {author} {\bibfnamefont
  {F.}~\bibnamefont {Boschini}}, \bibinfo {author} {\bibfnamefont
  {H.}~\bibnamefont {Chu}}, \bibinfo {author} {\bibfnamefont {Y.}~\bibnamefont
  {Murakami}}, \bibinfo {author} {\bibfnamefont {G.}~\bibnamefont {Levy}},
  \bibinfo {author} {\bibfnamefont {A.~K.}\ \bibnamefont {Mills}}, \bibinfo
  {author} {\bibfnamefont {S.}~\bibnamefont {Zhdanovich}}, \bibinfo {author}
  {\bibfnamefont {M.}~\bibnamefont {Isobe}}, \bibinfo {author} {\bibfnamefont
  {H.}~\bibnamefont {Takagi}}, \bibinfo {author} {\bibfnamefont
  {S.}~\bibnamefont {Kaiser}}, \bibinfo {author} {\bibfnamefont
  {P.}~\bibnamefont {Werner}}, \bibinfo {author} {\bibfnamefont {D.~J.}\
  \bibnamefont {Jones}}, \bibinfo {author} {\bibfnamefont {A.}~\bibnamefont
  {Georges}}, \bibinfo {author} {\bibfnamefont {A.}~\bibnamefont {Damascelli}},
  \ and\ \bibinfo {author} {\bibfnamefont {A.~J.}\ \bibnamefont {Millis}},\
  }\href {\doibase 10.1103/PhysRevB.106.L121106} {\bibfield  {journal}
  {\bibinfo  {journal} {Phys. Rev. B}\ }\textbf {\bibinfo {volume} {106}},\
  \bibinfo {pages} {L121106} (\bibinfo {year} {2022})}\BibitemShut {NoStop}%
\bibitem [{\citenamefont {Chen}\ \emph {et~al.}(2025)\citenamefont {Chen},
  \citenamefont {Mravlje}, \citenamefont {Golež},\ and\ \citenamefont
  {Werner}}]{chen20252dcoherentspectroscopysignatures}%
  \BibitemOpen
  \bibfield  {author} {\bibinfo {author} {\bibfnamefont {J.}~\bibnamefont
  {Chen}}, \bibinfo {author} {\bibfnamefont {J.}~\bibnamefont {Mravlje}},
  \bibinfo {author} {\bibfnamefont {D.}~\bibnamefont {Golež}}, \ and\ \bibinfo
  {author} {\bibfnamefont {P.}~\bibnamefont {Werner}},\ }\href
  {https://arxiv.org/abs/2512.19689} {\enquote {\bibinfo {title} {2d coherent
  spectroscopy signatures of exciton condensation in ta$_2$nise$_5$},}\ }
  (\bibinfo {year} {2025}),\ \Eprint {http://arxiv.org/abs/2512.19689}
  {arXiv:2512.19689 [cond-mat.str-el]} \BibitemShut {NoStop}%
\bibitem [{\citenamefont {Chatterjee}\ \emph {et~al.}(2025)\citenamefont
  {Chatterjee}, \citenamefont {Mravlje},\ and\ \citenamefont
  {Gole\ifmmode~\check{z}\else \v{z}\fi{}}}]{chatterjee_collective_modes_2025}%
  \BibitemOpen
  \bibfield  {author} {\bibinfo {author} {\bibfnamefont {B.}~\bibnamefont
  {Chatterjee}}, \bibinfo {author} {\bibfnamefont {J.}~\bibnamefont {Mravlje}},
  \ and\ \bibinfo {author} {\bibfnamefont {D.}~\bibnamefont
  {Gole\ifmmode~\check{z}\else \v{z}\fi{}}},\ }\href {\doibase
  10.1103/PhysRevB.111.L121106} {\bibfield  {journal} {\bibinfo  {journal}
  {Phys. Rev. B}\ }\textbf {\bibinfo {volume} {111}},\ \bibinfo {pages}
  {L121106} (\bibinfo {year} {2025})}\BibitemShut {NoStop}%
\bibitem [{\citenamefont {Xuan}\ \emph {et~al.}(2026)\citenamefont {Xuan},
  \citenamefont {Song},\ and\ \citenamefont
  {Sun}}]{Xuan_collective_modes_EI_2025}%
  \BibitemOpen
  \bibfield  {author} {\bibinfo {author} {\bibfnamefont {F.}~\bibnamefont
  {Xuan}}, \bibinfo {author} {\bibfnamefont {J.}~\bibnamefont {Song}}, \ and\
  \bibinfo {author} {\bibfnamefont {Z.}~\bibnamefont {Sun}},\ }\href {\doibase
  10.1103/rdrs-ndjc} {\bibfield  {journal} {\bibinfo  {journal} {Phys. Rev. B}\
  }\textbf {\bibinfo {volume} {113}},\ \bibinfo {pages} {035117} (\bibinfo
  {year} {2026})}\BibitemShut {NoStop}%
\bibitem [{\citenamefont {Katsumi}\ \emph {et~al.}(2023)\citenamefont
  {Katsumi}, \citenamefont {Alekhin}, \citenamefont {Souliou}, \citenamefont
  {Merz}, \citenamefont {Haghighirad}, \citenamefont {Le~Tacon}, \citenamefont
  {Houver}, \citenamefont {Cazayous}, \citenamefont {Sacuto},\ and\
  \citenamefont {Gallais}}]{katsumi_disentangling_PRL2023}%
  \BibitemOpen
  \bibfield  {author} {\bibinfo {author} {\bibfnamefont {K.}~\bibnamefont
  {Katsumi}}, \bibinfo {author} {\bibfnamefont {A.}~\bibnamefont {Alekhin}},
  \bibinfo {author} {\bibfnamefont {S.-M.}\ \bibnamefont {Souliou}}, \bibinfo
  {author} {\bibfnamefont {M.}~\bibnamefont {Merz}}, \bibinfo {author}
  {\bibfnamefont {A.-A.}\ \bibnamefont {Haghighirad}}, \bibinfo {author}
  {\bibfnamefont {M.}~\bibnamefont {Le~Tacon}}, \bibinfo {author}
  {\bibfnamefont {S.}~\bibnamefont {Houver}}, \bibinfo {author} {\bibfnamefont
  {M.}~\bibnamefont {Cazayous}}, \bibinfo {author} {\bibfnamefont
  {A.}~\bibnamefont {Sacuto}}, \ and\ \bibinfo {author} {\bibfnamefont
  {Y.}~\bibnamefont {Gallais}},\ }\href {\doibase
  10.1103/PhysRevLett.130.106904} {\bibfield  {journal} {\bibinfo  {journal}
  {Phys. Rev. Lett.}\ }\textbf {\bibinfo {volume} {130}},\ \bibinfo {pages}
  {106904} (\bibinfo {year} {2023})}\BibitemShut {NoStop}%
\bibitem [{\citenamefont {Baldini}\ \emph {et~al.}(2023)\citenamefont
  {Baldini}, \citenamefont {Zong}, \citenamefont {Choi}, \citenamefont {Lee},
  \citenamefont {Michael}, \citenamefont {Windgaetter}, \citenamefont {Mazin},
  \citenamefont {Latini}, \citenamefont {Azoury}, \citenamefont {Lv},
  \citenamefont {Kogar}, \citenamefont {Su}, \citenamefont {Wang},
  \citenamefont {Lu}, \citenamefont {Takayama}, \citenamefont {Takagi},
  \citenamefont {Millis}, \citenamefont {Rubio}, \citenamefont {Demler},\ and\
  \citenamefont {Gedik}}]{baldini_TNS_pnas2023}%
  \BibitemOpen
  \bibfield  {author} {\bibinfo {author} {\bibfnamefont {E.}~\bibnamefont
  {Baldini}}, \bibinfo {author} {\bibfnamefont {A.}~\bibnamefont {Zong}},
  \bibinfo {author} {\bibfnamefont {D.}~\bibnamefont {Choi}}, \bibinfo {author}
  {\bibfnamefont {C.}~\bibnamefont {Lee}}, \bibinfo {author} {\bibfnamefont
  {M.~H.}\ \bibnamefont {Michael}}, \bibinfo {author} {\bibfnamefont
  {L.}~\bibnamefont {Windgaetter}}, \bibinfo {author} {\bibfnamefont {I.~I.}\
  \bibnamefont {Mazin}}, \bibinfo {author} {\bibfnamefont {S.}~\bibnamefont
  {Latini}}, \bibinfo {author} {\bibfnamefont {D.}~\bibnamefont {Azoury}},
  \bibinfo {author} {\bibfnamefont {B.}~\bibnamefont {Lv}}, \bibinfo {author}
  {\bibfnamefont {A.}~\bibnamefont {Kogar}}, \bibinfo {author} {\bibfnamefont
  {Y.}~\bibnamefont {Su}}, \bibinfo {author} {\bibfnamefont {Y.}~\bibnamefont
  {Wang}}, \bibinfo {author} {\bibfnamefont {Y.}~\bibnamefont {Lu}}, \bibinfo
  {author} {\bibfnamefont {T.}~\bibnamefont {Takayama}}, \bibinfo {author}
  {\bibfnamefont {H.}~\bibnamefont {Takagi}}, \bibinfo {author} {\bibfnamefont
  {A.~J.}\ \bibnamefont {Millis}}, \bibinfo {author} {\bibfnamefont
  {A.}~\bibnamefont {Rubio}}, \bibinfo {author} {\bibfnamefont
  {E.}~\bibnamefont {Demler}}, \ and\ \bibinfo {author} {\bibfnamefont
  {N.}~\bibnamefont {Gedik}},\ }\href {\doibase 10.1073/pnas.2221688120}
  {\bibfield  {journal} {\bibinfo  {journal} {Proceedings of the National
  Academy of Sciences}\ }\textbf {\bibinfo {volume} {120}},\ \bibinfo {pages}
  {e2221688120} (\bibinfo {year} {2023})},\ \Eprint
  {http://arxiv.org/abs/https://www.pnas.org/doi/pdf/10.1073/pnas.2221688120}
  {https://www.pnas.org/doi/pdf/10.1073/pnas.2221688120} \BibitemShut {NoStop}%
\bibitem [{\citenamefont {Wei}\ \emph {et~al.}(2025)\citenamefont {Wei},
  \citenamefont {Luo}, \citenamefont {Watanabe}, \citenamefont {Taniguchi},
  \citenamefont {Guo},\ and\ \citenamefont {Xi}}]{wei_gate_tuningTNS}%
  \BibitemOpen
  \bibfield  {author} {\bibinfo {author} {\bibfnamefont {K.}~\bibnamefont
  {Wei}}, \bibinfo {author} {\bibfnamefont {Y.}~\bibnamefont {Luo}}, \bibinfo
  {author} {\bibfnamefont {K.}~\bibnamefont {Watanabe}}, \bibinfo {author}
  {\bibfnamefont {T.}~\bibnamefont {Taniguchi}}, \bibinfo {author}
  {\bibfnamefont {Y.}~\bibnamefont {Guo}}, \ and\ \bibinfo {author}
  {\bibfnamefont {X.}~\bibnamefont {Xi}},\ }\href {\doibase
  10.1038/s41467-025-66594-y} {\bibfield  {journal} {\bibinfo  {journal}
  {Nature Communications}\ }\textbf {\bibinfo {volume} {16}},\ \bibinfo {pages}
  {10999} (\bibinfo {year} {2025})}\BibitemShut {NoStop}%
\bibitem [{\citenamefont {Shi}\ \emph {et~al.}(2025)\citenamefont {Shi},
  \citenamefont {Zhang}, \citenamefont {Huang}, \citenamefont {Isobe},
  \citenamefont {Takagi}, \citenamefont {Keimer},\ and\ \citenamefont
  {Boris}}]{uniaxial_pressure_control}%
  \BibitemOpen
  \bibfield  {author} {\bibinfo {author} {\bibfnamefont {X.}~\bibnamefont
  {Shi}}, \bibinfo {author} {\bibfnamefont {Y.-S.}\ \bibnamefont {Zhang}},
  \bibinfo {author} {\bibfnamefont {D.}~\bibnamefont {Huang}}, \bibinfo
  {author} {\bibfnamefont {M.}~\bibnamefont {Isobe}}, \bibinfo {author}
  {\bibfnamefont {H.}~\bibnamefont {Takagi}}, \bibinfo {author} {\bibfnamefont
  {B.}~\bibnamefont {Keimer}}, \ and\ \bibinfo {author} {\bibfnamefont {A.~V.}\
  \bibnamefont {Boris}},\ }\href {\doibase 10.1103/jysr-2dk1} {\bibfield
  {journal} {\bibinfo  {journal} {Phys. Rev. Lett.}\ }\textbf {\bibinfo
  {volume} {135}},\ \bibinfo {pages} {156503} (\bibinfo {year}
  {2025})}\BibitemShut {NoStop}%
\bibitem [{\citenamefont {Rosenberg}\ \emph {et~al.}(2026)\citenamefont
  {Rosenberg}, \citenamefont {Ayres-Sims}, \citenamefont {Millis},
  \citenamefont {Cobden},\ and\ \citenamefont
  {Chu}}]{rosenberg2025elastocaloricsignatureexcitonicinstability}%
  \BibitemOpen
  \bibfield  {author} {\bibinfo {author} {\bibfnamefont {E.}~\bibnamefont
  {Rosenberg}}, \bibinfo {author} {\bibfnamefont {J.}~\bibnamefont
  {Ayres-Sims}}, \bibinfo {author} {\bibfnamefont {A.}~\bibnamefont {Millis}},
  \bibinfo {author} {\bibfnamefont {D.}~\bibnamefont {Cobden}}, \ and\ \bibinfo
  {author} {\bibfnamefont {J.-H.}\ \bibnamefont {Chu}},\ }\href {\doibase
  10.1103/sgdx-15kn} {\bibfield  {journal} {\bibinfo  {journal} {Phys. Rev.
  Lett.}\ }\textbf {\bibinfo {volume} {136}},\ \bibinfo {pages} {046503}
  (\bibinfo {year} {2026})}\BibitemShut {NoStop}%
\bibitem [{\citenamefont {Bae}\ \emph {et~al.}(2025)\citenamefont {Bae},
  \citenamefont {Raghavan}, \citenamefont {Feldman}, \citenamefont {Kanigel},\
  and\ \citenamefont {Madhavan}}]{bae2025microscopicevidencedominantexcitonic}%
  \BibitemOpen
  \bibfield  {author} {\bibinfo {author} {\bibfnamefont {S.}~\bibnamefont
  {Bae}}, \bibinfo {author} {\bibfnamefont {A.}~\bibnamefont {Raghavan}},
  \bibinfo {author} {\bibfnamefont {I.}~\bibnamefont {Feldman}}, \bibinfo
  {author} {\bibfnamefont {A.}~\bibnamefont {Kanigel}}, \ and\ \bibinfo
  {author} {\bibfnamefont {V.}~\bibnamefont {Madhavan}},\ }\href
  {https://arxiv.org/abs/2512.03011} {\enquote {\bibinfo {title} {Microscopic
  evidence of dominant excitonic instability in ta2nise5},}\ } (\bibinfo {year}
  {2025}),\ \Eprint {http://arxiv.org/abs/2512.03011} {arXiv:2512.03011
  [cond-mat.str-el]} \BibitemShut {NoStop}%
\bibitem [{\citenamefont {Sugimoto}\ \emph {et~al.}(2018)\citenamefont
  {Sugimoto}, \citenamefont {Nishimoto}, \citenamefont {Kaneko},\ and\
  \citenamefont {Ohta}}]{sugimoto_strong_couplingTNS}%
  \BibitemOpen
  \bibfield  {author} {\bibinfo {author} {\bibfnamefont {K.}~\bibnamefont
  {Sugimoto}}, \bibinfo {author} {\bibfnamefont {S.}~\bibnamefont {Nishimoto}},
  \bibinfo {author} {\bibfnamefont {T.}~\bibnamefont {Kaneko}}, \ and\ \bibinfo
  {author} {\bibfnamefont {Y.}~\bibnamefont {Ohta}},\ }\href {\doibase
  10.1103/PhysRevLett.120.247602} {\bibfield  {journal} {\bibinfo  {journal}
  {Phys. Rev. Lett.}\ }\textbf {\bibinfo {volume} {120}},\ \bibinfo {pages}
  {247602} (\bibinfo {year} {2018})}\BibitemShut {NoStop}%
\bibitem [{sup()}]{suppl}%
  \BibitemOpen
  \href@noop {} {}\bibinfo {note} {See Supplemental Material containing
  additional details on the model Hamiltonian, the charge and current
  distributions of the symmetry broken phases, and the lattice
  dynamics.}\BibitemShut {Stop}%
\bibitem [{\citenamefont {Peierls}(1933)}]{peierls_original}%
  \BibitemOpen
  \bibfield  {author} {\bibinfo {author} {\bibfnamefont {R.}~\bibnamefont
  {Peierls}},\ }\href {\doibase 10.1007/BF01342591} {\bibfield  {journal}
  {\bibinfo  {journal} {Zeitschrift f{\"u}r Physik}\ }\textbf {\bibinfo
  {volume} {80}},\ \bibinfo {pages} {763} (\bibinfo {year} {1933})}\BibitemShut
  {NoStop}%
\bibitem [{\citenamefont {Luttinger}(1951)}]{luttinger_peierls_phases1951}%
  \BibitemOpen
  \bibfield  {author} {\bibinfo {author} {\bibfnamefont {J.~M.}\ \bibnamefont
  {Luttinger}},\ }\href {\doibase 10.1103/PhysRev.84.814} {\bibfield  {journal}
  {\bibinfo  {journal} {Phys. Rev.}\ }\textbf {\bibinfo {volume} {84}},\
  \bibinfo {pages} {814} (\bibinfo {year} {1951})}\BibitemShut {NoStop}%
\bibitem [{\citenamefont {Li}\ \emph {et~al.}(2020)\citenamefont {Li},
  \citenamefont {Golez}, \citenamefont {Mazza}, \citenamefont {Millis},
  \citenamefont {Georges},\ and\ \citenamefont
  {Eckstein}}]{electromagnetic_coupling_2020}%
  \BibitemOpen
  \bibfield  {author} {\bibinfo {author} {\bibfnamefont {J.}~\bibnamefont
  {Li}}, \bibinfo {author} {\bibfnamefont {D.}~\bibnamefont {Golez}}, \bibinfo
  {author} {\bibfnamefont {G.}~\bibnamefont {Mazza}}, \bibinfo {author}
  {\bibfnamefont {A.~J.}\ \bibnamefont {Millis}}, \bibinfo {author}
  {\bibfnamefont {A.}~\bibnamefont {Georges}}, \ and\ \bibinfo {author}
  {\bibfnamefont {M.}~\bibnamefont {Eckstein}},\ }\href {\doibase
  10.1103/PhysRevB.101.205140} {\bibfield  {journal} {\bibinfo  {journal}
  {Phys. Rev. B}\ }\textbf {\bibinfo {volume} {101}},\ \bibinfo {pages}
  {205140} (\bibinfo {year} {2020})}\BibitemShut {NoStop}%
\bibitem [{\citenamefont {Matsuda}\ \emph {et~al.}(2020)\citenamefont
  {Matsuda}, \citenamefont {Nakamura}, \citenamefont {Ikeda}, \citenamefont
  {Takeyama}, \citenamefont {Suga}, \citenamefont {Nakahara},\ and\
  \citenamefont {Muraoka}}]{magnetic_fild_V02_ncomm2020}%
  \BibitemOpen
  \bibfield  {author} {\bibinfo {author} {\bibfnamefont {Y.~H.}\ \bibnamefont
  {Matsuda}}, \bibinfo {author} {\bibfnamefont {D.}~\bibnamefont {Nakamura}},
  \bibinfo {author} {\bibfnamefont {A.}~\bibnamefont {Ikeda}}, \bibinfo
  {author} {\bibfnamefont {S.}~\bibnamefont {Takeyama}}, \bibinfo {author}
  {\bibfnamefont {Y.}~\bibnamefont {Suga}}, \bibinfo {author} {\bibfnamefont
  {H.}~\bibnamefont {Nakahara}}, \ and\ \bibinfo {author} {\bibfnamefont
  {Y.}~\bibnamefont {Muraoka}},\ }\href {\doibase 10.1038/s41467-020-17416-w}
  {\bibfield  {journal} {\bibinfo  {journal} {Nature Communications}\ }\textbf
  {\bibinfo {volume} {11}},\ \bibinfo {pages} {3591} (\bibinfo {year}
  {2020})}\BibitemShut {NoStop}%
\bibitem [{\citenamefont {Fernandes}\ \emph {et~al.}(2026)\citenamefont
  {Fernandes}, \citenamefont {Birol}, \citenamefont {Ye},\ and\ \citenamefont
  {Vanderbilt}}]{loop_current_kagome}%
  \BibitemOpen
  \bibfield  {author} {\bibinfo {author} {\bibfnamefont {R.~M.}\ \bibnamefont
  {Fernandes}}, \bibinfo {author} {\bibfnamefont {T.}~\bibnamefont {Birol}},
  \bibinfo {author} {\bibfnamefont {M.}~\bibnamefont {Ye}}, \ and\ \bibinfo
  {author} {\bibfnamefont {D.}~\bibnamefont {Vanderbilt}},\ }\href {\doibase
  10.1038/s41567-026-03229-z} {\bibfield  {journal} {\bibinfo  {journal}
  {Nature Physics}\ }\textbf {\bibinfo {volume} {22}},\ \bibinfo {pages} {991}
  (\bibinfo {year} {2026})}\BibitemShut {NoStop}%
\bibitem [{\citenamefont {Cioni}\ \emph {et~al.}(2026)\citenamefont {Cioni},
  \citenamefont {Cavicchi}, \citenamefont {Africani}, \citenamefont {Mazza},
  \citenamefont {Taddei}, \citenamefont {Yacoby},\ and\ \citenamefont
  {Polini}}]{cioni2026_persistent_currents}%
  \BibitemOpen
  \bibfield  {author} {\bibinfo {author} {\bibfnamefont {F.}~\bibnamefont
  {Cioni}}, \bibinfo {author} {\bibfnamefont {L.}~\bibnamefont {Cavicchi}},
  \bibinfo {author} {\bibfnamefont {N.}~\bibnamefont {Africani}}, \bibinfo
  {author} {\bibfnamefont {G.}~\bibnamefont {Mazza}}, \bibinfo {author}
  {\bibfnamefont {F.}~\bibnamefont {Taddei}}, \bibinfo {author} {\bibfnamefont
  {A.}~\bibnamefont {Yacoby}}, \ and\ \bibinfo {author} {\bibfnamefont
  {M.}~\bibnamefont {Polini}},\ }\href {https://arxiv.org/abs/2606.09812}
  {\enquote {\bibinfo {title} {Persistent currents, whirlpools, and local chern
  markers in twisted tmd chern insulators},}\ } (\bibinfo {year} {2026}),\
  \Eprint {http://arxiv.org/abs/2606.09812} {arXiv:2606.09812
  [cond-mat.mes-hall]} \BibitemShut {NoStop}%
\bibitem [{\citenamefont {Mazza}(2026)}]{mazza_2026_22232200}%
  \BibitemOpen
  \bibfield  {author} {\bibinfo {author} {\bibfnamefont {G.}~\bibnamefont
  {Mazza}},\ }\href {\doibase 10.5281/zenodo.22232200} {\enquote {\bibinfo
  {title} {Magnetic field control of the excitonic transition in ta2nise5},}\ }
  (\bibinfo {year} {2026})\BibitemShut {NoStop}%
\end{thebibliography}%

\end{document}


\author{Giacomo Mazza \\
{\it Dipartimento di Fisica dell'Universit\`a di Pisa, Largo Bruno Pontecorvo 3, I-56127 Pisa,~Italy}}

\title{
Supplmental Information: \\
Magnetic field control of the excitonic transition in \tns
}

\maketitle

\onecolumngrid
\tableofcontents

\section{Model Hamiltonian and mean-field decoupling}
\begin{figure}
\includegraphics[width=0.7\columnwidth]{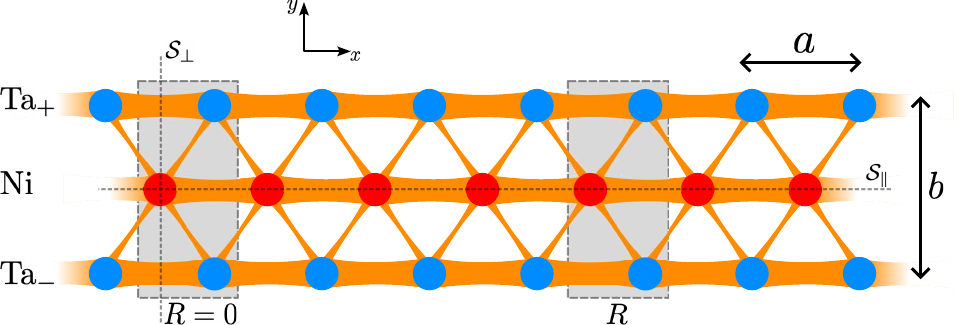}
\caption{Chain structure of \tns in the orhtorombic phase.}
\end{figure}

The electronic Hamiltonian considered in the paper is
\begin{equation}
H_{el} = H_0 + H_{int}
\end{equation}
where $H_0$ is the tight-binding Hamiltonian and 
$H_{int}$ contains density-density interaction terms.
The tight-binding Hamiltonian is built from Wannier 
orbitals localized on the upper and lower Ta-atoms 
and on the central Ni-atom, and it reads
\begin{equation}
H_0 = \sum_{\s} \sum_{\a \b }^{\left\lbrace \tpm,\nik\right\rbrace} 
\sum_{R R'} t_{R,R'}^{\a \b} 
\cc_{R \a \s} \ca_{R' \b \s}.
\end{equation}
We recall the symmetries of the \tns tight-binding Hamiltonian
referring the reader to Ref.~\cite{giacomo_TNS} for an in-depth discussion. 
The Ta- and Ni- Wannier orbitals are, respectively, 
even and odd with respect to the reflections about the 
the transverse plane ${\cal S}_{\perp}$.
Specifically, by indicating with $(x,y)$ the spatial 
coordinates of the chain plane with origin $x=0$ and $y=0$
on the Ni-atom of the $R=0$ cell, and ${\cal S}_{\perp}$ the $x=0$ 
plane, the Wannier orbitals centred on the atom $\a$ of the cell 
$R$, $w_{\alpha, R} (x,y)$, transform under reflections
about the ${\cal S}_{\perp}$ plane accordingly to
\begin{equation}
\begin{split}
&w_{R, \tpm} (x,y) = w_{-R-a, \tpm}(-x,y) \\
&w_{R, \nik} (x,y) = -w_{-R,\nik}(-x,y).
\end{split}
\end{equation}
By considering reflections about the $y=0$ plane $({\cal S}_{\parallel})$
\begin{equation}
\begin{split}
&w_{R,\tpm } (x,y) = w_{R,\tmp}(x,-y) \\
&w_{R,\nik } (x,y) = w_{ R, \nik}(x,-y).
\end{split}
\end{equation} 
This reflection symmetries constraint the hopping parameters
describing the $\tpm-\nik$ hybridization as 
\begin{equation}
t_{R,R}^{\tpm \nik} = -t_{R,R-a}^{\tpm \nik}.
\end{equation}  
Analogously, we define the trasnformation of the fermionic operators under the reflection 
symmetries ${\cal R}_{\cal S_{\perp}}$ and ${\cal R}_{\cal S_{\parallel}}$ as 
\begin{equation}
\begin{split}
&{\cal R}_{\cal S_\perp} \cc_{R,\tpm \s} \to  \cc_{-R-a,\tpm \s} \\
&{\cal R}_{\cal S_\perp} \cc_{R,\nik \s} \to  -\cc_{-R,\nik \s},
\end{split}
\end{equation}
and 
\begin{equation}
\begin{split}
&{\cal R}_{\cal S_\parallel} \cc_{R,\tpm \s} \to  \cc_{R,\tmp \s} \\
&{\cal R}_{\cal S_\parallel} \cc_{R,\nik \s} \to  \cc_{R,\nik \s}.
\end{split}
\end{equation}
It follows that, due to the invariance under ${\cal R}_{\cal S_{\perp}}$ 
the $\G-$point hybridization identically vanishes
\begin{equation}
\Psi_{\pm} = \sum_{\s} \quave{\cc_{k=0,\tpm,\s} \ca_{k=0,\nik,\s}}
= \frac{1}{N^2} \sum_{R, R'} \quave{\cc_{R,\tpm,\s} \ca_{R',\nik,\s}} 
= \frac{1}{N} \sum_{R} \quave{\cc_{R,\tpm,\s} \ca_{0,\nik,\s}} 
\end{equation}
\begin{equation}
\Psi_{\pm} = {\cal R}_{\cal S_\perp} \Psi_{\pm} = 
- \frac{1}{N} \sum_{R} \quave{\cc_{-R-a,\tpm,\s} \ca_{0,\nik,\s}} = 
- \Psi_{\pm} = 0.
\end{equation} 
Moreover, invariance under ${\cal R}_{\cal S_\parallel}$ implies $\Psi_{+} = \Psi_-$.

The Hamiltonian $H_{int}$ contains intra- and inter-chain 
density-density interaction terms 
\begin{equation}
H_{int} = \frac{1}{2}\sum_{\s \s'} \sum_{\a \b} \sum_{R R'} 
U_{\a \b}^{\s \s'} (R-R') \cc_{R \a \s} \ca_{R \a \s} \cc_{R' \b \s'} \ca_{R' \b \s'}.
\end{equation}
We consider long-range interaction potentials decaying as $|R|^{-1}$, 
and parametrized by the intracell interaction strengths 
\begin{equation}
U_{\a \a}^{\s \s'}(R) = U_\a \frac{a}{|R|+a} \left[ \d_{\s,\s'} \left( 1-  \d_{R,0}\right) + \d_{\s -\s'} \right],
\end{equation}
and 
\begin{equation}
U_{\tpm,\nik}^{\s \s'} (R) = V \frac{a}{|R| + |R+a|}.
\end{equation}
We neglect $\tplus-\tmins$ interaction $U_{\tplus, \tmins} = 0$.

We solve the interacting model by considering the Hartree-Fock 
mean-field decoupling in the particle-hole channel imposing the 
translational invariance of the chain. 
We define the particle-hole amplitudes 
\begin{align}
\Delta_{\a \b}^{\s} (k) \equiv \quave{\cc_{k \a \s} \ca_{k \b \s}}  
\end{align}
which, upon decoupling of the interaction, are 
self-consistently determined by the ground state 
of the Hamiltonian
\begin{align}
H_{HF} = H_0 + \sum_{\a \b} \sum_{\s} 
\cc_{k \a \s} \ca_{k \b \s}
W^\s_{\a \b}(k) + {\rm h.c.}
\end{align}
where the effect of the interaction is 
included in the self-consistent fields 
\begin{align}
W^\s_{\a \b}(k) \equiv 
- \sum_{k'} U_{\a \b}^{\s \s} (k-k') \left[ \Delta_{\a \b}^{\s} (k') \right]^*.
\label{eq:Wself}
\end{align}
Notice that the divergent of the Hartree term 
is absorbed into a redefinition of the chemical potential to fix 
the average number of electron per unit cell, and is not included in the self-consistent Hamiltonian.
To carry out the momentum integration in Eq.~\ref{eq:Wself}, we smear out the $q \to 0$ singularity of the interaction by 
introducing a tiny exponential cut-off, and 
check convergence with respect to the latter.

\section{Charge and current density }
In this section, we analyse the charge and current density in the CD and 
LC excitonic phases discussed in the main text.
\subsection*{Charge Density}
The charge density reads
\begin{align}
\rho(x,y) = \sum_{\a \b} \sum_{R R'} \sum_{\s} 
w_{R,\a}(x,y) w_{R', \b}(x,y) \quave{\cc_{R \a \s} \ca_{R' \b \s}}
=
\sum_{\a} \rho_{\a \a}(x,y) + \sum_{\a \neq \b} \rho_{\a \neq \b} (x,y)
\end{align}
where $\rho_{\a \a} (x,y) $ 
are the diagonal component. The off-diagonal components further split into
$\tplus-\tmins$ and $\tpm-\nik$ contributions as 
$\rho_{\a \neq \b} (x,y) = \rho_{\rm \tplus,\tmins} (x,y) + \rho_{\rm {\rm Ta},\nik} (x,y)$
with  
\begin{align}
&\rho_{\a,\a}(x,y) = \frac{1}{2}
\sum_{R,R'} \sum_{\s} 
w_{R,\a}(x,y) w_{R', \a}(x,y) 
\left[ \quave{\cc_{R \a \s} \ca_{R' \a \s}} + c.c. \right] \\
&\rho_{\tplus,\tmins}(x,y) = \sum_{R,R'} \sum_{\s} 
w_{R,\tplus}(x,y) w_{R', \tmins}(x,y) \left[ \quave{\cc_{R \tplus \s} \ca_{R' \tmins \s}} + c.c. \right]\\
&\rho_{{\rm Ta},\nik}(x,y) = \sum_{\a=\tpm} \sum_{R,R'} \sum_{\s} 
w_{R,\a}(x,y) w_{R', \nik}(x,y) \left[ \quave{\cc_{R \a \s} \ca_{R' \nik \s}} + c.c. \right]
\end{align}
We analyse the symmetry with respect to the reflections. 
Let us first consider the symmetry with respect to ${\cal S}_\perp$, and 
and compute the charge asymmetry
\begin{align}
\d_{\perp}(x,y) = \rho(x,y) - \rho(-x,y).
\end{align}
 with
\begin{align}
\rho(-x,y) = \sum_{\a} \rho_{\a \a} (-x,y) + \rho_{\tplus,\tmins}(-x,y) + \rho_{{\rm Ta},\nik} (-x,y).
\end{align}
By exploiting the translational invariance, i.e. 
$\quave{\ca_{R \a \s} \ca_{R' \b \s}} = \quave{\ca_{R+R'' \a \s} \ca_{R'+R'' \b \s}}$ 
for any $R''$, we get
\begin{itemize}
\item[-] diagonal component
\begin{align}
\rho_{\a \a} (x,y) - \rho_{\a \a} (-x,y) = \sum_{R R'} \sum_{\s} w_{R \a}(x,y) 
w_{R' \a}(x,y) {\rm Re}\left[ \quave{\cc_{(R-R') \a \s} \ca_{0 \a \s}} - \quave{\cc_{(R'-R) \a \s} \ca_{0 \a \s}} \right]
\end{align}
\item[-] off-diagonal components
\begin{align}
& \rho_{\tplus \tmins} (x,y) - \rho_{\tplus \tmins} (-x,y) = 
\sum_{R R'} \sum_{\s} w_{R \tplus}(x,y) 
w_{R' \tmins}(x,y) 
{\rm Re} \left[ 
\quave{\cc_{(R-R') \tplus \s} \ca_{0 \tmins \s}} - 
\quave{\cc_{(R'-R) \tplus \s} \ca_{0 \tmins \s}} \right]\\
&\rho_{{\rm Ta} \nik} (x,y) - \rho_{{\rm Ta} \nik} (-x,y) =
\sum_{\a =\tpm} 
\sum_{R R'} \sum_\s
w_{R \a}(x,y) w_{R' \nik} (x,y)
{\rm Re} 
\left[  
\quave{\cc_{(R-R') \a \s} \ca_{0 \nik \s}} + 
\quave{\cc_{(R'-R -a) \a \s} \ca_{0 \nik \s}} 
\right]
\label{eq:off_diag_TN_cd_perp}
\end{align}
\end{itemize}
It follows that the charge is symmetric 
with respect to reflections about the $x=0$ plane 
if the intra- and inter-orbital amplitudes obey
\begin{equation}
\begin{split}
&{\rm Re} \quave{\cc_{R \a \s} \ca_{0 \a \s}} = {\rm Re} \quave{\cc_{-R \a \s} \ca_{0 \a \s}} \\
&{\rm Re} \quave{\cc_{R \tplus \s} \ca_{0 \tmins \s}} = 
{\rm Re} \quave{\cc_{-R \tplus \s} \ca_{0 \tmins \s}}
\qquad \qquad \longrightarrow \qquad \qquad \d_{\perp}(x,y) = 0 \\
&{\rm Re} \quave{\cc_{R \tpm \s} \ca_{0 \nik \s}} = 
-{\rm Re} \quave{\cc_{-R-a \tpm \s} \ca_{0 \nik \s}} \\
\end{split}
\end{equation}

Let us consider now the reflection with respect to the longitudinal plane
\begin{equation}
\d_{\parallel} (x,y) = \rho(x,y) - \rho(x,-y)
\end{equation}
\begin{itemize}
\item[-] diagonal components
\begin{equation}
\begin{split}
&\sum_{\a=\tpm}
\rho_{\a\a}(x,y)-\rho_{\a\a} (x,-y)
= \sum_{R R'} 
\left[ w_{R \tplus} (x,y) w_{R' \tplus} (x,y) - 
w_{R \tmins} (x,y) w_{R' \tmins} (x,y) \right] \times \\
&\phantom{\rho_{\tpm \tpm}(x,y)-\rho_{\tpm \tpm} (x,-y)= \qquad \qquad} 
\times {\rm Re}\left[ \quave{\cc_{(R-R') \tpm \s} \ca_{0 \tpm \s} } - 
\quave{\cc_{(R-R') \tmp \s} \ca_{0 \tmp \s} } \right] \\
&\rho_{\nik \nik}(x,y)-\rho_{\nik \nik} (x,-y) = 0
\end{split}
\end{equation}
\item[-] off-diagonal components
\begin{align}
&
\rho_{\tplus \tmins}(x,y)-\rho_{\tplus \tmins} (x,-y)
= 
\sum_{R R'} w_{R \tplus} (x,y) w_{R' \tmins} (x,y)
{\rm Re} \left[ \quave{\cc_{(R-R') \tplus \s} \ca_{0 \tmins \s}} - 
\quave{\cc_{-(R-R') \tmp \s} \ca_{0 \tpm \s}} \right] \\
&\rho_{\tpm \nik}(x,y)-\rho_{\tpm \nik} (x,-y)
= 
\sum_{R R'} w_{R \tpm} (x,y) w_{R' \tmp} (x,y)
{\rm Re}
\left[ 
\quave{\cc_{(R-R') \tpm \s} \ca_{0 \nik \s}} - \quave{\cc_{(R-R') \tmp \s} \ca_{0 \nik \s} }
\right]
\label{eq:off_diag_TN_cd_parallel}
\end{align}
\end{itemize}

It follows that the charge is symmetric 
with respect to reflections about the $y=0$ plane 
if the intra- and inter-orbital amplitudes obey
\begin{equation}
\begin{split}
&{\rm Re} \quave{\cc_{R \tpm \s} \ca_{0 \tpm \s}} = {\rm Re} \quave{\cc_{R \tmp \s} \ca_{0 \tmp \s}} \\
&{\rm Re} \quave{\cc_{R \tplus \s} \ca_{0 \tmins \s}} = 
{\rm Re} \quave{\cc_{-R \tplus \s} \ca_{0 \tmins \s}}
\qquad \qquad \longrightarrow \qquad \qquad \d_{\parallel}(x,y) = 0 \\
&{\rm Re} \quave{\cc_{R \tpm \s} \ca_{0 \nik \s}} = 
{\rm Re} \quave{\cc_{R  \tmp \s} \ca_{0 \nik \s}} 
\\
\end{split}
\end{equation}

In Fig.~\ref{fig:supp_fig1} (first row) we show the real parts of the 
diagonal and off-diagonal 
amplitudes $\D_{\s}^{\a \b} (R) 
\equiv \quave{\cc_{R \a \s} \ca_{0 \b \s}}$ for the 
CD $\varphi=0$ and LC $\varphi = \pi/2$ excitonic insulating  phases.
We see that for the CD phase the charge density 
is not invariant under both reflection symmetries, i. $\d_{\perp} \neq 0$ and $\d_{\parallel} \neq 0$.
In particular, symmetry breaking is evident in the 
off-diagonal components, $(\tplus,\tmins)$ and $(\tpm,\nik)$. 
We further notice that the 
$(\tplus,\tmins)$ component of the charge density 
transforms in the same way under both symmetries.
Moreover, we see that the $(\tplus,\nik)$ and $(\tmins,\nik)$ 
components are related by
\begin{equation}
\D_{\s}^{\tmins,\nik}(R) = - \D_{\s}^{\tplus,\nik}(-R-a)
\end{equation}
so that the conditions \ref{eq:off_diag_TN_cd_perp} and \ref{eq:off_diag_TN_cd_parallel}
becomes equivalent, meaning that also the $(\tplus,\nik)$ and $(\tmins,\nik)$ 
components transform in the same way under both symmetry.
As a result, charge density remains invariant under the simultaneous 
action of the two reflections, leading to the charge density distribution for 
the CD phase sketched in the main text.

By looking at the same quantities for the LC phase, it is immediate to check 
that, for this phase, the charge density preserves both reflection symmetries.

\subsection*{Current Density}
We now consider the current distribution.
The current density operator 
is defined by taking the variation of the Hamiltonian with 
respect to a probe field  $\vec{A}(\vec{r})$ 
\begin{equation}
\vec{j}(\vec{r}) = - \left. \frac{\d H(\vec{A})}{\d \vec{A}(\vec{r})} \right|_{\vec{A}=0}
\end{equation}
Since we are considering a tight-binding model, the probe field 
enter through Peierls phases 
\begin{equation}
H(\vec{A}) = \sum_{R R'} \sum_{\a \b} t_{R R'}^{\a \b} 
\exp\left[- i \frac{e}{\hbar} 
\int^{\vec{r}_{R \a}}_{\vec{r}_{R' \b}} d \vec{r} \cdot \vec{A}(\vec{r}) \right]
\cc_{R \a} \ca_{R' \b} + H_{int}
\end{equation}
where  $\int^{\vec{r}_{R \a}}_{\vec{r}_{R' \b}} d \vec{r} \cdot \vec{A}(\vec{r}) $
is a line integral along a straight line connecting the 
Wannier centers $R' \b$ and $R \a$. To proceed, it is custom to assume that 
the probe field varies smoothly along the integration line, so that 
one can approximate
\begin{align*}
\int^{\vec{r}_{R \a}}_{\vec{r}_{R' \b}} d \vec{r} \cdot \vec{A}(\vec{r})  
\simeq \vec{A}_{R \a, R' \b} \cdot \left[ \vec{r}_{R \a} -\vec{r}_{R' \b} \right]
\end{align*}
where $\vec{A}_{R \a, R' \b}$ is the vector potential evaluated at the 
mid point $\vec{r}_{R \a, R' \b} = \frac{\vec{r}_{R \a} + \vec{r}_{R' \b}}{2}$ of the straight line. 
In doing so, the functional derivative leads to a current density 
defined on the mid-point of links between lattice sites as 
\begin{equation}
\vec{j}(\vec{r}) = -\frac{e}{\hbar} \sum_{R R'} \sum_{\a \b} 
\d(\vec{r}-\vec{r}_{R \a, R' \b}) 
\left[ \vec{r}_{R \a} -\vec{r}_{R' \b} \right] 2 {\rm Im } 
\left( t_{R R'}^{\a \b} \quave{\cc_{R \a} \ca_{R' \b}} \right) =
\sum_{R R'} \sum_{\a \b} \d(\vec{r}-\vec{r}_{R \a, R' \b}) 
\vec{u}_{R \a, R' \b}
{I}_{R R'}^{\a \b}
\end{equation}
where $\vec{u}_{R \a, R' \b} = \frac{\vec{r}_{R \a} -\vec{r}_{R' \b}
}{\left| \vec{r}_{R \a} -\vec{r}_{R' \b} \right|} $ 
is the link unit vector, and ${I}_{R R'}^{\a \b}$ is the intensity 
current on the link. 
Since the intensity current proportional to the hopping, 
only nearest-neighboring links have non zero current.
In the following, we list the expressions for the 
currents amplitude for the nearest neighboring links in the absence 
of external fields for which the hopping amplitudes are real. The summation over 
spin is understood.
\begin{align}
& \vec{j}_{R,R-a}^{\tpm \tpm} 
= -\frac{e}{\hbar} 
\hat{x} a 2 t^{\tpm \tpm}_{R, R-a} {\rm Im} (\quave{\cc_{a \tpm \s } \ca_{0 \tpm \s}}) \\
& 
\vec{j}_{R,R-a}^{\nik \nik} 
= -\frac{e}{\hbar} 
\hat{x} a 2 t^{\nik \nik}_{R, R-a} {\rm Im} (\quave{\cc_{a \nik \s } \ca_{0 \nik \s}}) \\
& 
\vec{j}_{R,R}^{\tpm \nik} = 
-\frac{e}{\hbar}  
\left( - a \hat{x}  - b \hat{y} \right)
t^{\tpm,\nik}_{R,R} {\rm Im} (\quave{\cc_{0 \tpm \s } \ca_{0 \nik \s}}) \\
& 
\vec{j}_{R,R-a}^{\tpm \nik} = 
-\frac{e}{\hbar} \left(a \hat{x} - b \hat{y} \right)
t^{\tpm,\nik}_{R,R-a} {\rm Im} (\quave{\cc_{a \tpm \s } \ca_{0 \nik \s}}) 
\end{align}
where $\hat{x}$ and $\hat{y}$ are the $x,y$ unit vectors and 
$a$ and $b$ are the lattice parameters.

\begin{figure}[h]
\includegraphics[width=\textwidth]{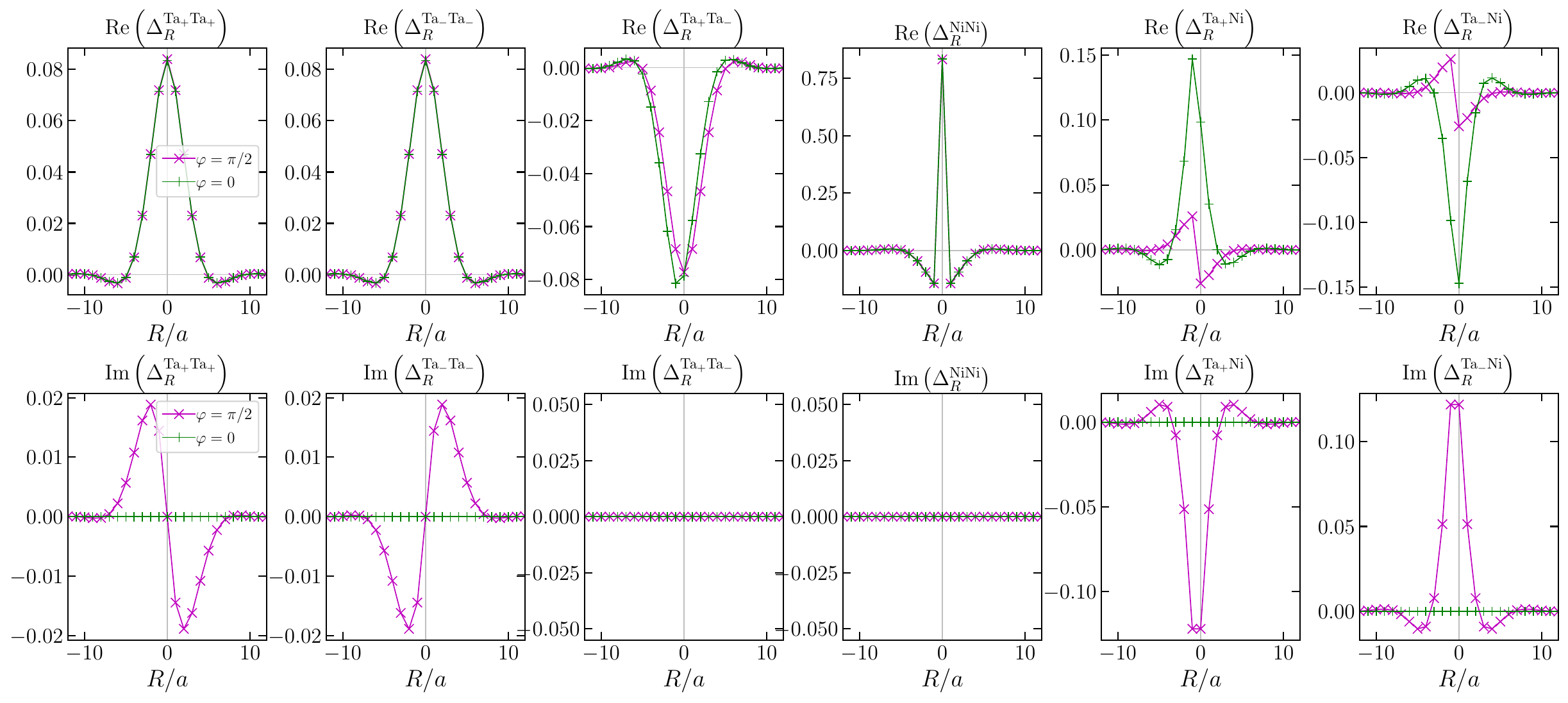}
\caption{Real (first row) and imaginary (second row) parts
of the amplitudes $\D_{\s}^{\a \b} (R) \equiv \quave{\cc_{R \a \s} \ca_{0 \b \s}}$
in the CD $(\varphi=0)$, green plus symbols, and LC $(\varphi=\pi/2)$ phase, magenta 
cross symbols.}
\label{fig:supp_fig1}
\end{figure}

In Fig.~\ref{fig:supp_fig1} (second row) we show the imaginary parts 
of the amplitudes $\D_{\s}^{\a \b} (R)$. In the CD phase, 
all the amplitudes are purely real so that the ground 
state current is always zero.
On the contrary, in the LC phase the imaginary parts leads to 
a pattern of currents which, by considering the signs of the 
hoppings, coincides with the pattern sketched in the main text.


\newpage
\section{\added{Stability of the excitonic phases}}

In this section, we show the details of the stability 
analysis of the CD and LC excitonic phases.

In order to check the stability of the LC and CD phases we use
a contrained Hartree-Fock minimisation to compute the energy 
as a function of an input order parameter.
To this extent we introduce Lagrange parameters $\L$ and minimize 
the variational energy at fixed value of excitonic order parameter
\begin{equation}
E[\Psi] = \braket{\Phi_{HF}}{H}{\Phi_{HF}} + \Lambda\left( \Psi - 
\braket{\Phi_{HF}}{\hat{\Psi}}{\Phi_{HF}} \right) + c.c.,
\label{eq:Epsi}
\end{equation}
where $\hat{\Psi} = \sum_{R \s} \sum_{\a=\pm} \alpha \l (t) \left[
\cc_{R \tplus \sigma} \ca_{R \nik \s} + \cc_{R \tplus \sigma} \ca_{R-1 \nik \s} \right]$ is the operator associated with the excitonic order parameter.
The self-consistent LC and solutions corresponding to the CD 
and LC phases correspond to stationary points of the function 
in Eq.~\ref{eq:Epsi}. By analyzing the energy near the 
stationary point we check if the stationary solution is a 
stable/metastable phase, namely a global/local minimum, or 
an unstable phase, namely a saddle point.

Additionally, we double check the stability of the phases 
using a time-dependent fluctuation analysis.
Starting from the CD/LC stationary solutions, we add a 
tiny delta-like perturbation that couples with the 
real/imaginary parts of the order parameter as
\begin{equation}
H(t) = H + \sum_{R \s} \sum_{\a=\pm} \alpha \l (t) \left[
\cc_{R \tplus \sigma} \ca_{R \nik \s} + \cc_{R \tplus \sigma} \ca_{R-1 \nik \s} \right] +
{\rm h.c.}.
\end{equation} 
with $\lambda(t) = \l e^{i \varphi} \exp(-(t-t_0)^2/2\t^2)$, 
where $\varphi = 0$ and $\varphi=\pi/2$ if we are checking 
the stability of the LC and CD phase, respectively.
We choose  $\t=1~{\rm fs}$ much smaller than the characteristic 
period of the order parameter oscillations and study the dynamics 
using time-dependent Hartree-Fock for different values of $\l$. 
If the stationary solution is stable minimum, we expect the 
order parameter dynamics to remain confined around the 
stationary solution with variation of proportional to the 
applied perturbation.
Otherwise, the dynamics will flow away from the stationary 
solution independently of the strength of the perturbation.

In subsection~\ref{sec:fluct_el} we analyze stability in the purely electronic model. 
In subsection~\ref{sec:fluct_eph}, we analyze stability with respect
to the electron-phonon coupling at zero applied field, and 
in subsection~\ref{sec:fluct_eph_field} we include also 
the applied field.

\subsection{Purely electronic model}
\label{sec:fluct_el}
\begin{figure}
\includegraphics[width=0.6\textwidth]{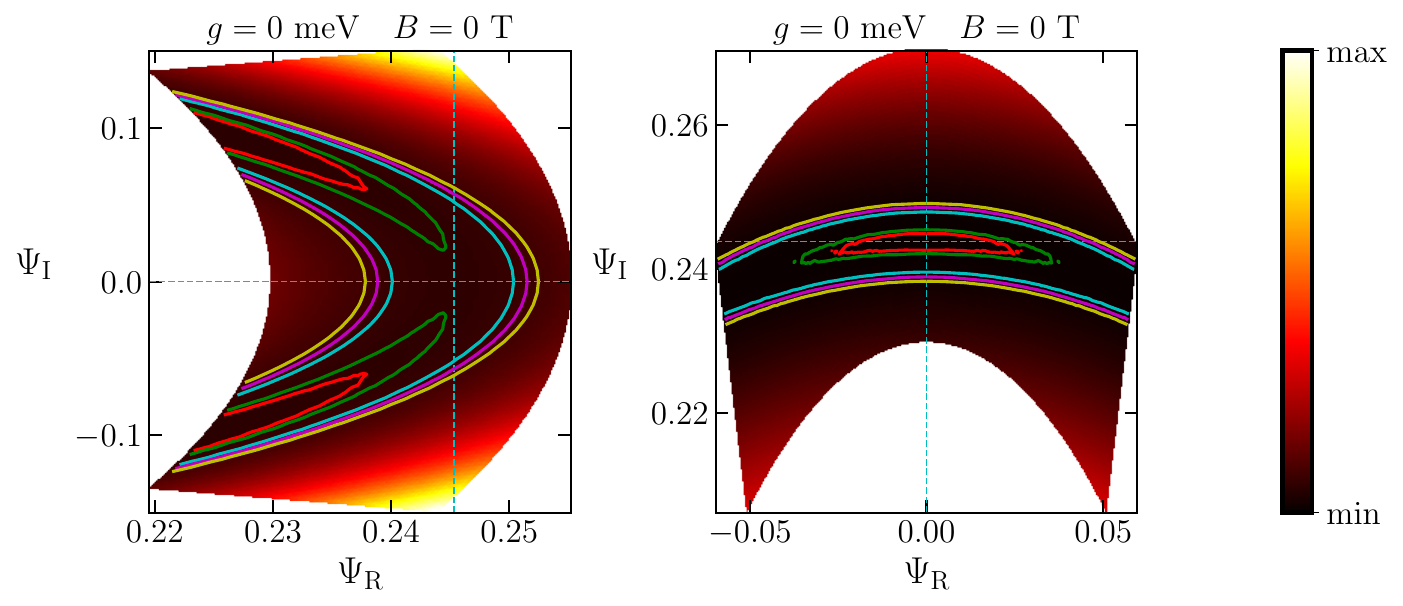}
\caption{ 
Energy as a function of the order parameter 
for $g=0~{\rm meV}$ and $B=0~{\rm T}$.
Model parameter $V=1.1~{\rm eV}.$
(Left)  energy landscape near the CD stationary point. 
(Right) energy landscape near the LC stationary point. 
Contour lines at increasing values of the energy: red, green, cyan, magenta, and yellow
from smallest to largest values of the energy.
Dashed lines highlight the stationary solutions.
}
\label{fig:energy_VS_OP_g0_B0}
\end{figure}

\begin{figure}
\includegraphics[width=0.8\textwidth]{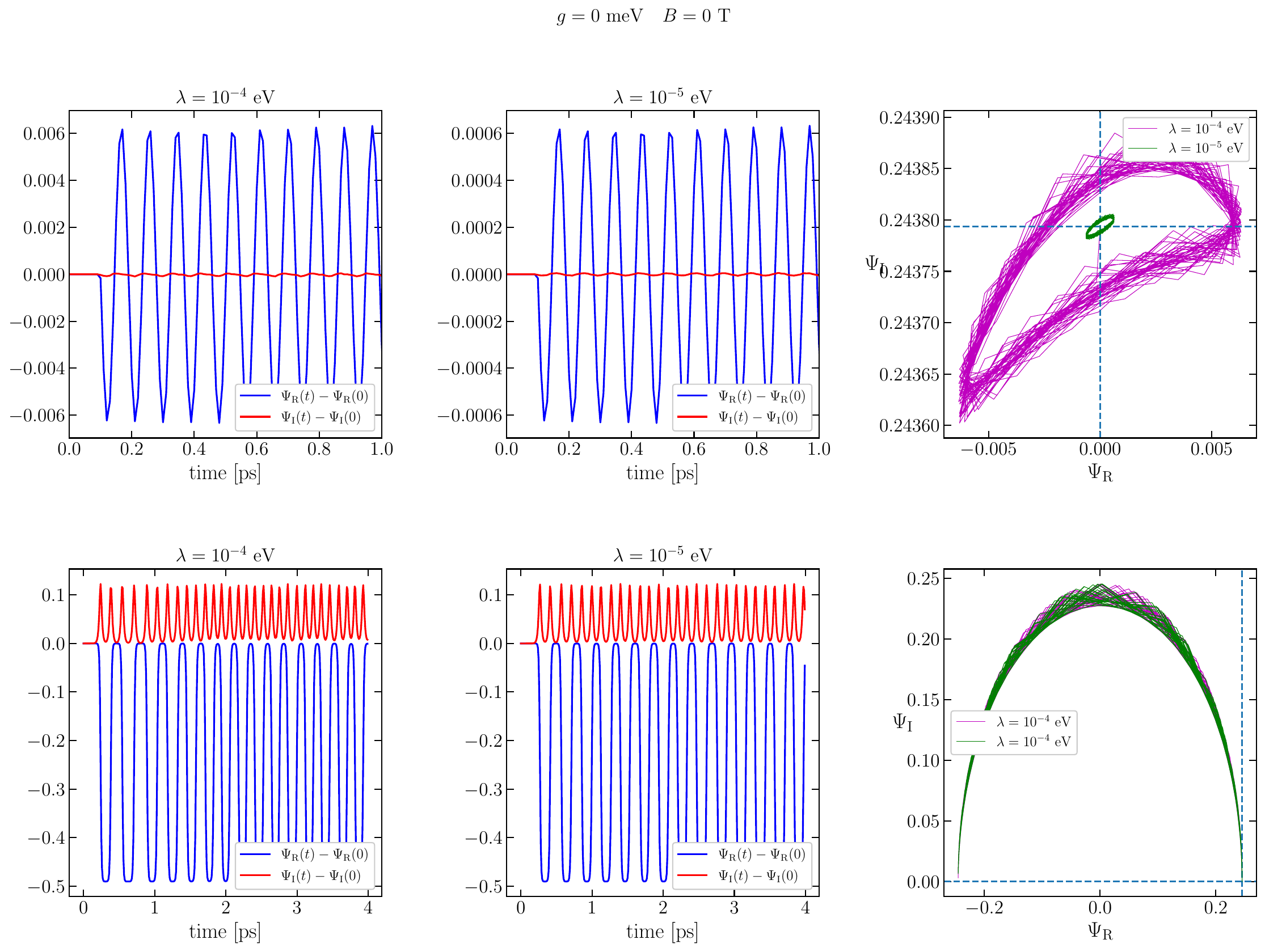}
\caption{
Time-dependent fluctuation analysis for $g=0~{\rm meV}$ and $B=0~{\rm T}$.
Model parameter $V=1.1~{\rm eV}.$
(Left and middle panels): 
Dynamics of the real and imaginary parts of the excitonic 
order parameter,
$\Delta \Psi_{\rm I/R} = \Psi_{\rm I/R}(t)-\Psi_{\rm I/R}(0)$, 
for different values of the perturbation $\lambda=10^{-4}~{\rm eV}$ (left panel)
and $\lambda=10^{-5}~{\rm eV}$ (middle panel). 
(right panel) Dynamics in the $\Psi_{\rm R}-\Psi_{\rm I}$ 
space. 
Top/bottom rows correspond to the fluctuations around the LC/CD 
stationary solutions. 
Dashed lines highlight the stationary solutions.
}
\label{fig:fluctuation_g0_B0}
\end{figure}

We start by the purely electronic model, 
by setting $g=0~{\rm meV}$ and $B=0~{\rm T}.$  
In Fig.~\ref{fig:energy_VS_OP_g0_B0} we show 
the energy as a function of the order parameter 
around the CD and LC stationary solutions, 
$\Psi_{\rm R} \neq 0$ and $\Psi_{\rm I} = 0$ (left), 
and 
$\Psi_{\rm R} = 0$ and $\Psi_{\rm I} \neq 0$ (right), respectively.
From the contour lines we see that the LC phase is the 
only stable solution.
We confirm the stability of the LC phase by carrying 
out the time-dependent fluctuation analysis in Fig.~\ref{fig:fluctuation_g0_B0}.

\subsubsection{Stability of the excitonic phases in the double-chains model with short range interactions}
\begin{figure}
\includegraphics[width=0.6\textwidth]{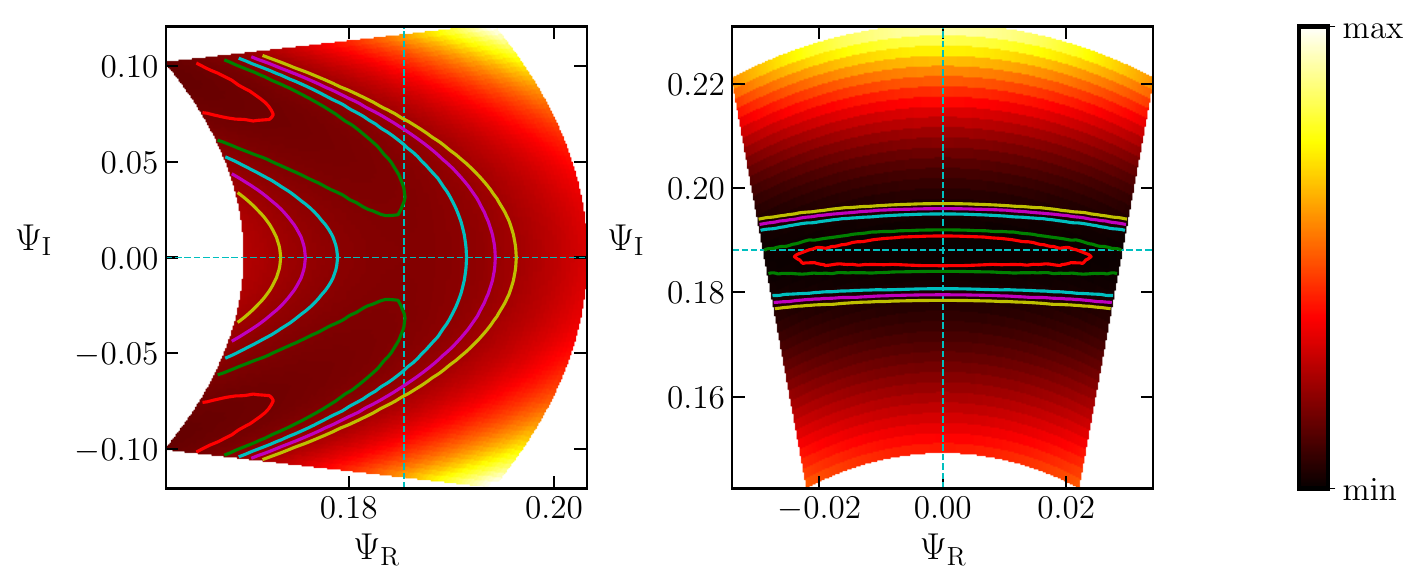}
\caption{Energy as a function of the order parameter in the double-chains
model with short range interactions, see Ref.~\cite{giacomo_TNS}.
(Left)  energy landscape near the CD stationary point. 
(Right) energy landscape near the LC stationary point. 
Contour lines at increasing values of the energy: red, green, cyan, magenta, and yellow
from smallest to largest values of the energy.
Dashed lines highlight the stationary solutions.}
\label{fig:energy_VS_OP_g0_B0_PRL2020}
\end{figure}

For completeness, we show that the LC phase is the actual 
stable excitonic phase also in the double chain model 
with short range interaction, originally introduced in Ref~\cite{giacomo_TNS} 
to illustrate the nature of the symmetry breaking of the excitonic phase in \tnscomma.
In Fig.~\ref{fig:energy_VS_OP_g0_B0_PRL2020} we show the energy 
as a function of the order parameter around the LC and CD stationary 
solutions using the same parameters of the bare bands of Ref.~\cite{giacomo_TNS} 
and the short range interaction parameters $U=2.50~{\rm eV}$ 
and $V=0.77~{\rm eV}$, inside the excitonic region of the phase diagram.

\subsection{Stability with electron-phonon coupling at zero-field}
\label{sec:fluct_eph}
We now consider the full Hamiltonian including the electron-phonon
coupling
\begin{equation}
H = H_{el} + H_{ph} + H_{ph-el}.
\end{equation}
We use the standard mean-field decoupling 
$\ket{\Psi} = \ket{\Psi_{el}} \ket{\Psi_{ph}}$ 
which leads to a set of coupled Schr\"odinger 
equations for the electronic and phononic wave functions
\begin{align}
&\tilde{H}_{el} \ket{\Psi_{el}} = E_{el} \ket{\Psi_{el}} \\
&\tilde{H}_{ph} \ket{\Psi_{ph}} = E_{ph} \ket{\Psi_{ph}} 
\end{align}
where the effective Hamiltonians read
\begin{align*}
\tilde{H}_{ph} = H_{ph} + \braket{\Psi_{el}}{H_{el-ph}}{\Psi_{el}} 
\qquad{\rm and} \qquad
\tilde{H}_{el} = H_{el} + \braket{\Psi_{ph}}{H_{el-ph}}{\Psi_{ph}}.
\end{align*}
We notice that the effective electronic Hamiltonian depends 
on the electron-phonon coupling only through the static distortion 
of the mode $\quave{X_{R \a}}$
\begin{align*}
\tilde{H}_{el} = H_{el} + g 
\sum_{\a} 
\quave{X_{R \a}} \left( \cc_{R \nik \s}  \ca_{R \a \s} + \cc_{R \nik \s}  \ca_{R-a \a \s} + {\rm h.c.} \right).
\end{align*}
It follows that, as mentioned in the main text, 
within the mean-field decoupling, any undistorted 
solution of the electron-phonon problem is 
equivalent to the $g=0$ case.

\begin{figure}
\includegraphics[width=0.6\textwidth]{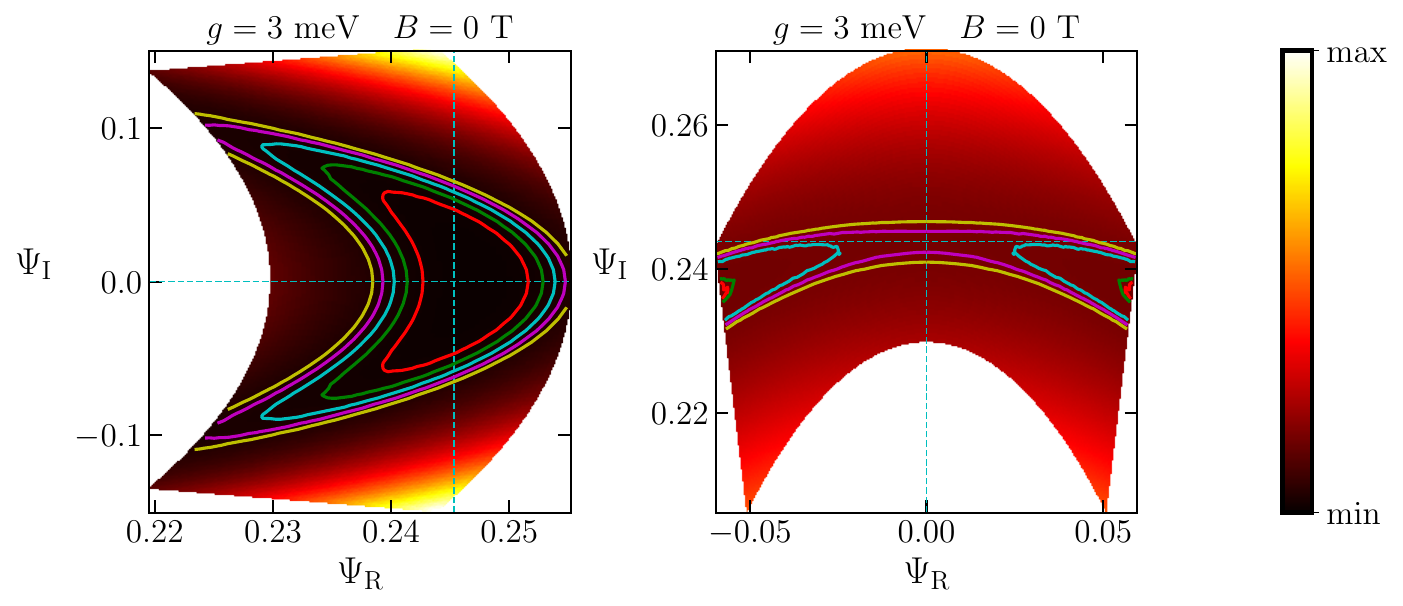}
\caption{ 
Energy as a function of the order parameter 
for $g=0.003~{\rm meV}$ and $B=0~{\rm T}$.
Model parameter $V=1.1~{\rm eV}.$
(Left)  energy landscape near the CD stationary point. 
(Right) energy landscape near the LC stationary point. 
Contour lines at increasing values of the energy: red, green, cyan, magenta, and yellow
from smallest to largest values of the energy.
Dashed lines highlight the stationary solutions.
}
\label{fig:energy_VS_OP_g0.003_B0}
\end{figure}

\begin{figure}
\includegraphics[width=0.9\textwidth]{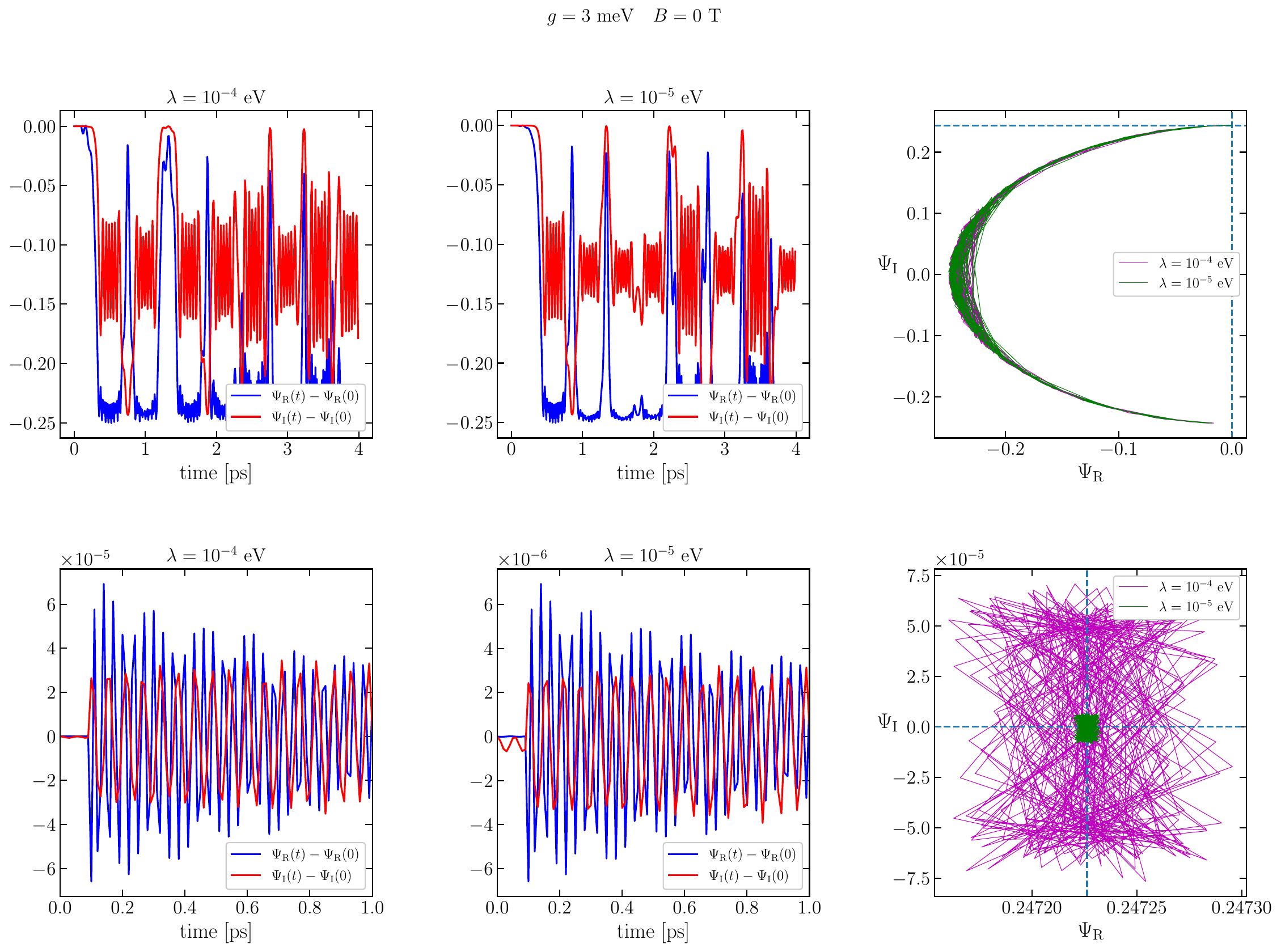}
\caption{
Time-dependent fluctuation analysis for $g=0.003~{\rm meV}$ and $B=0~{\rm T}$.
Model parameter $V=1.1~{\rm eV}.$
(Left and middle panels): 
Dynamics of the real and imaginary parts of the excitonic 
order parameter,
$\Delta \Psi_{\rm I/R} = \Psi_{\rm I/R}(t)-\Psi_{\rm I/R}(0)$, 
for different values of the perturbation $\lambda=10^{-4}~{\rm eV}$ (left panel)
and $\lambda=10^{-5}~{\rm eV}$ (middle panel). 
(right panel) Dynamics in the $\Psi_{\rm R}-\Psi_{\rm I}$ 
space. 
Top/bottom rows correspond to the fluctuations around the LC/CD 
stationary solutions. 
Dashed lines highlight the stationary solutions.
}
\label{fig:fluctuation_g0.003_B0}
\end{figure}

In Fig.~\ref{fig:energy_VS_OP_g0.003_B0} we show 
the energy as a function of the order parameter 
around the CD and LC stationary solutions, 
$\Psi_{\rm R} \neq 0$ and $\Psi_{\rm I} = 0$ (left), 
and 
$\Psi_{\rm R} = 0$ and $\Psi_{\rm I} \neq 0$ (right), respectively,
for $g=3~{\rm meV}$.
The CD saddle point for $g=0~{\rm meV}$ clearly 
evolves into the global minimum for $g=3~{\rm meV}$.
At the same time, the LC minimum for $g=0~{\rm meV}$ 
becomes a saddle point for $g=3~{\rm meV}$.
The switch of stability of the two phases 
occurs for the value of $g$ at which the 
energies of the two phases cross, see main text.
As done before, we confirm the stability switching by 
time-dependent fluctuation analysis in Fig.~\ref{fig:fluctuation_g0.003_B0}.

\subsection{Stability with electron-phonon coupling at finite field}
\label{sec:fluct_eph_field}
\begin{figure}
\includegraphics[width=0.6\textwidth]{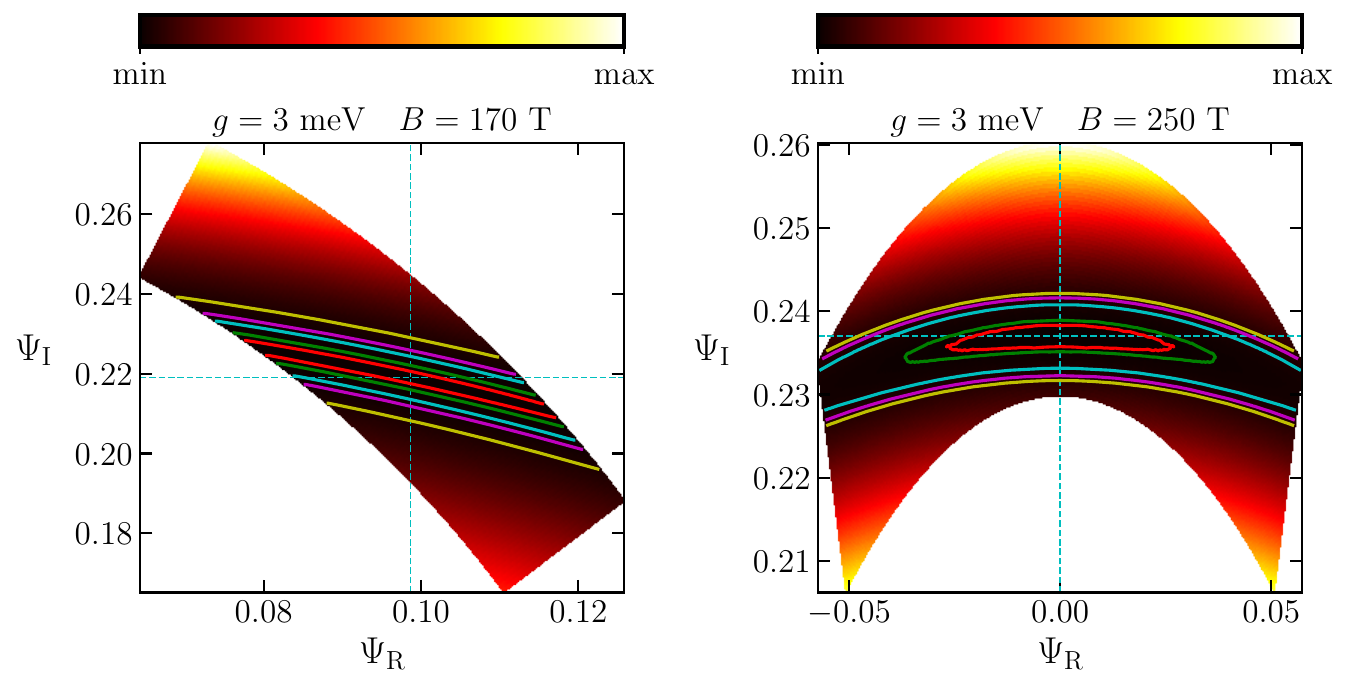}
\caption{ 
Energy as a function of the order parameter 
for $g=0.003~{\rm meV}$ and $B=170~{\rm T} < B_c$ (right) and 
$B=250~{\rm T} > B_c$.
Model parameter $V=1.1~{\rm eV}.$
Contour lines at increasing values of the energy: red, green, cyan, magenta, and yellow
from smallest to largest values of the energy.
Dashed lines highlight the stationary solutions.
}
\label{fig:energy_vs_op_g0.003_Bfield}
\end{figure}

Eventually, we show the recovery of the stability 
of the LC phase upon application of the field.
Increasing the applied field, the stable minimum continuously 
moves from the purely real to the purely imaginary order parameter.
In Fig.~\ref{fig:energy_vs_op_g0.003_Bfield}, we show the energy landscape 
for $g=3~{\rm meV}$ for two values of the applied field, 
before and after the transition occurring for $B_c \approx 190~{\rm T}$.

\section{Lattice dynamics}
\begin{figure}
\includegraphics[width=0.7\textwidth]{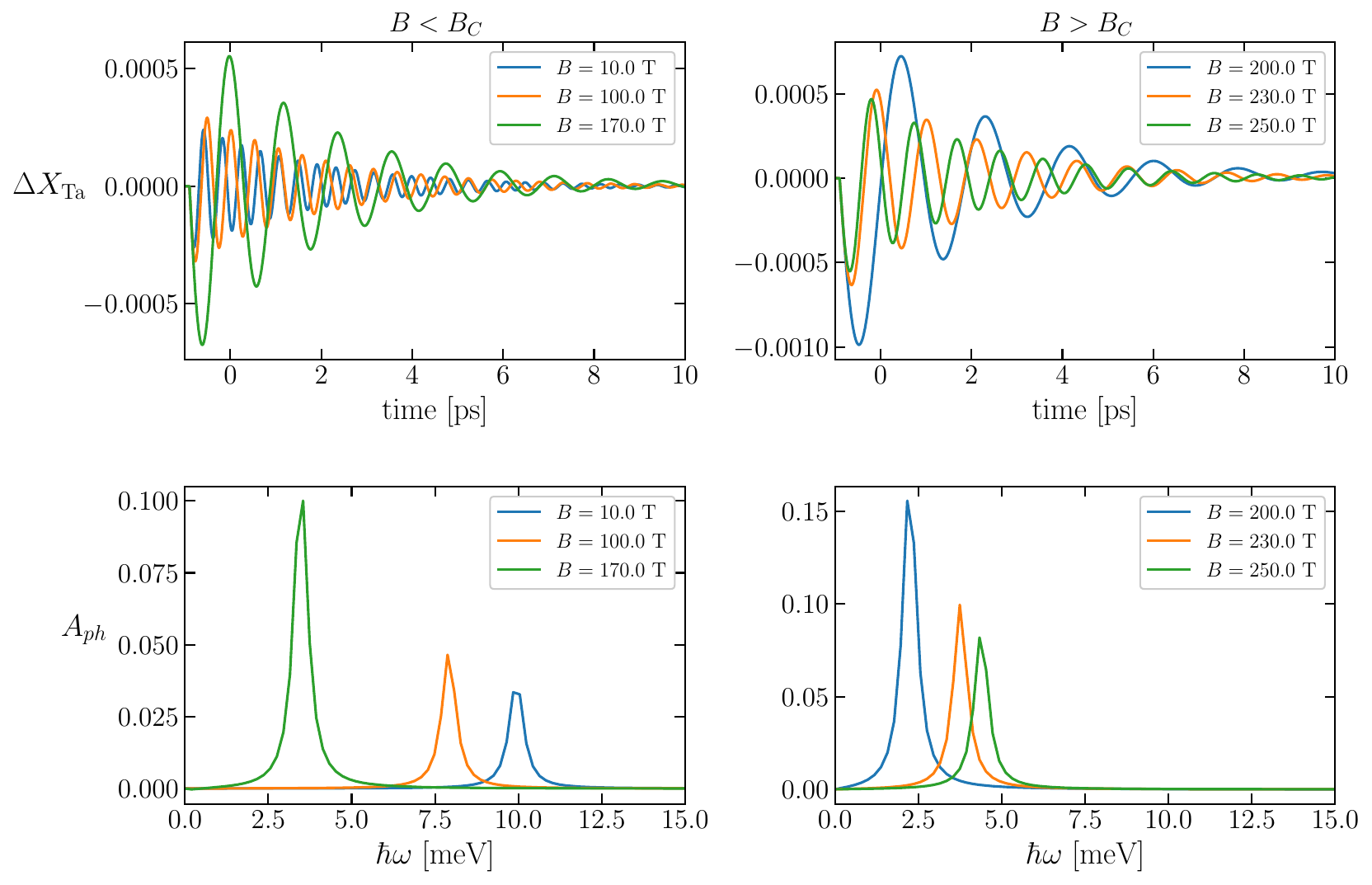}
\caption{(First row) Dynamics of the shear mode displacement
for different values of the applied fields in the structurally 
distorted (left) and the undistorted LC phase (right) for $g=3~{\rm meV}$.
(Second row) Corresponding phonon spectral functions.}
\label{fig:lattice_dynamics}
\end{figure}

In this section, we describe the lattice dynamics used to determine 
the shear mode frequency.
We add a time-dependent perturbation that couples to the 
phonon displacement 
\begin{align}
H \to H(t) = H + \delta(t) \sum_{R} (X_{R \tplus} - X_{R \tmins})
\label{eq:Hpht}
\end{align}
with $\d(t) = \d \exp(-(t-t_0)^2/2\t^2)$, and study the dynamics 
with time-dependent Hartree-Fock.
For all stable solutions, we track the oscillation in time of the phonon displacement $\quave{X_{\rm Ta}}(t)$, and use linear response theory 
to express the mode oscillation to the phonon response function as 
\begin{align}
\quave{X_{\rm Ta}}(t) = \quave{X_{\rm Ta}}_0 + \int_{-\infty}^{+\infty} dt' \chi_{ph} (t-t') \l(t').
\label{eq:linear_response}
\end{align}
From Eq.~\ref{eq:linear_response}, we extract the response 
function $\chi_{ph}(t-t')$, and compute its Fourier transform.
Eventually, we determine the frequency of the mode by averaging the frequency 
over the spectral function $A_{ph}(\o) = -\frac{1}{\pi} {\rm Im} \chi_{ph}(\o)$
\begin{align}
\o_{ph} = \frac{\int d \o \o A_{ph}(\o)}{\int d \o qA_{ph}(\o)}
\end{align}
In Fig.\ref{fig:lattice_dynamics}, we show few examples of lattice dynamics 
across the magnetic-field induced monoclinic-to-orthorombic distortion.

\bibliography{biblio_TNS}

\newpage
\section{Figures}

\newpage

\newpage

\newpage


\newpage

